\documentclass[twocolumn, amsmath, amssymb, longbibliography,superscriptaddress]{revtex4-1} 
\pdfoutput=1

\usepackage{graphicx}
\usepackage{dcolumn}
\usepackage{bm}
\usepackage{bbm}
\usepackage{color}
\usepackage{url}
\usepackage{multirow} 
\usepackage[normalem]{ulem}
\usepackage{xcolor}
\usepackage{makecell}
\newcommand{\be}{\begin{equation}}
\newcommand{\ee}{\end{equation}}
\newcommand{\bd}{\begin{displaymath}}
\newcommand{\ed}{\end{displaymath}}
\newcommand{\BE}{\begin{eqnarray}}
\newcommand{\EE}{\end{eqnarray}}

\newcommand{\bra}{\langle}
\newcommand{\ket}{\rangle}

\newcommand{\bc}{\ensuremath{\mathbf{c}}}

\newcommand{\bs}{\ensuremath{\mathbf{s}}}

\newcommand{\bz}{\ensuremath{\mathbf{z}}}

\newcommand{\mcD}{\mathcal{D}}

\definecolor{darkgreen}{rgb}{0.0, 0.5, 0.0}
\DeclareMathOperator{\sign}{sign}

\begin{document}

\preprint{}

\title{Quantum-Assisted Learning of Hardware-Embedded Probabilistic Graphical Models}

\author{Marcello Benedetti}
\affiliation{Quantum Artificial Intelligence Laboratory, NASA Ames Research Center, Moffett Field, California 94035, USA}
\affiliation{Department of Computer Science, University College London, WC1E 6BT London, United Kingdom}
\affiliation{Cambridge Quantum Computing Limited, CB2 1UB Cambridge, United Kingdom}

\author{John Realpe-G\'omez}
\affiliation{Quantum Artificial Intelligence Laboratory, NASA Ames Research Center, Moffett Field, California 94035, USA}
\affiliation{SGT Inc., Greenbelt, Maryland 20770, USA}
\affiliation{Instituto de Matem\'aticas Aplicadas, Universidad de Cartagena, Bol\'ivar 130001, Colombia}

\author{Rupak Biswas}
\affiliation{Quantum Artificial Intelligence Laboratory, NASA Ames Research Center, Moffett Field, California 94035, USA}
\affiliation{Exploration Technology Directorate, NASA Ames Research Center, Moffett Field, California 94035, USA}

\author{Alejandro Perdomo-Ortiz}
\email{Correspondance: alejandro.perdomoortiz@nasa.gov}
\affiliation{Quantum Artificial Intelligence Laboratory, NASA Ames Research Center, Moffett Field, California 94035, USA}
\affiliation{Department of Computer Science, University College London, WC1E 6BT London, United Kingdom}
\affiliation{Cambridge Quantum Computing Limited, CB2 1UB Cambridge, United Kingdom}
\affiliation{USRA Research Institute for Advanced Computer Science, Mountain View, California 94043, USA}

\begin{abstract}
Mainstream machine-learning techniques such as deep learning and probabilistic programming rely heavily on sampling from generally intractable probability distributions. There is increasing interest in the potential advantages of using quantum computing technologies as sampling engines to speed up these tasks or to make them more effective. However, some pressing challenges in state-of-the-art quantum annealers have to be overcome before we can assess their actual performance. The sparse connectivity, resulting from the local interaction between quantum bits in physical hardware implementations, is considered the most severe limitation to the quality of constructing powerful generative unsupervised machine-learning models. Here we use embedding techniques to add redundancy to data sets, allowing us to increase the modeling capacity of quantum annealers. We illustrate our findings by training hardware-embedded graphical models on a binarized data set of handwritten digits and two synthetic data sets in experiments with up to $940$ quantum bits. Our model can be trained in quantum hardware without full knowledge of the effective parameters specifying the corresponding quantum Gibbs-like distribution; therefore, this approach avoids the need to infer the effective temperature at each iteration, speeding up learning; it also mitigates the effect of noise in the control parameters, making it robust to deviations from the reference Gibbs distribution. Our approach demonstrates the feasibility of using quantum annealers for implementing generative models, and it provides a suitable framework for benchmarking these quantum technologies on machine-learning-related tasks. 
\end{abstract}

\maketitle

\section{Introduction}
Sampling from high-dimensional probability distributions is at the core of a wide spectrum of machine-learning techniques with important applications across science, engineering, and society; deep learning~\cite{LeCun-Nature-2015} and probabilistic programming~\cite{Zoubin-Nature-2015} are some notable examples. While much of the record-breaking performance of machine-learning algorithms regularly reported in the literature pertains to task-specific supervised learning algorithms~\cite{LeCun-Nature-2015,Bengio-Book}, the development of the more humanlike unsupervised learning algorithms has been lagging behind. An approach to unsupervised learning is to model the joint probability distribution of all the variables of interest. This is known as the generative approach because it allows us to generate synthetic data by sampling from the joint distribution. Generative models find application in anomaly detection, reinforcement learning, handling of missing values, and visual arts, to name a few \cite{goodfellow2016nips}. Even in some supervised contexts, it may be useful to treat the targets as standard input and attempt to model the joint distribution~\cite{ng2002discriminative}. Generative models rely on a sampling engine that is used for both inference and learning. Because of the intractability of traditional sampling techniques like the Markov chain Monte Carlo (MCMC) method, finding good generative models is among the hardest problems in machine learning~\cite{LeCun-Nature-2015,Bengio-Book,Salakhutdinov-Review-2015,Sinclair-InfComp-1989,Frigessi-Biometrika-1997}. 

\begin{figure}
\includegraphics[width=.50\textwidth]{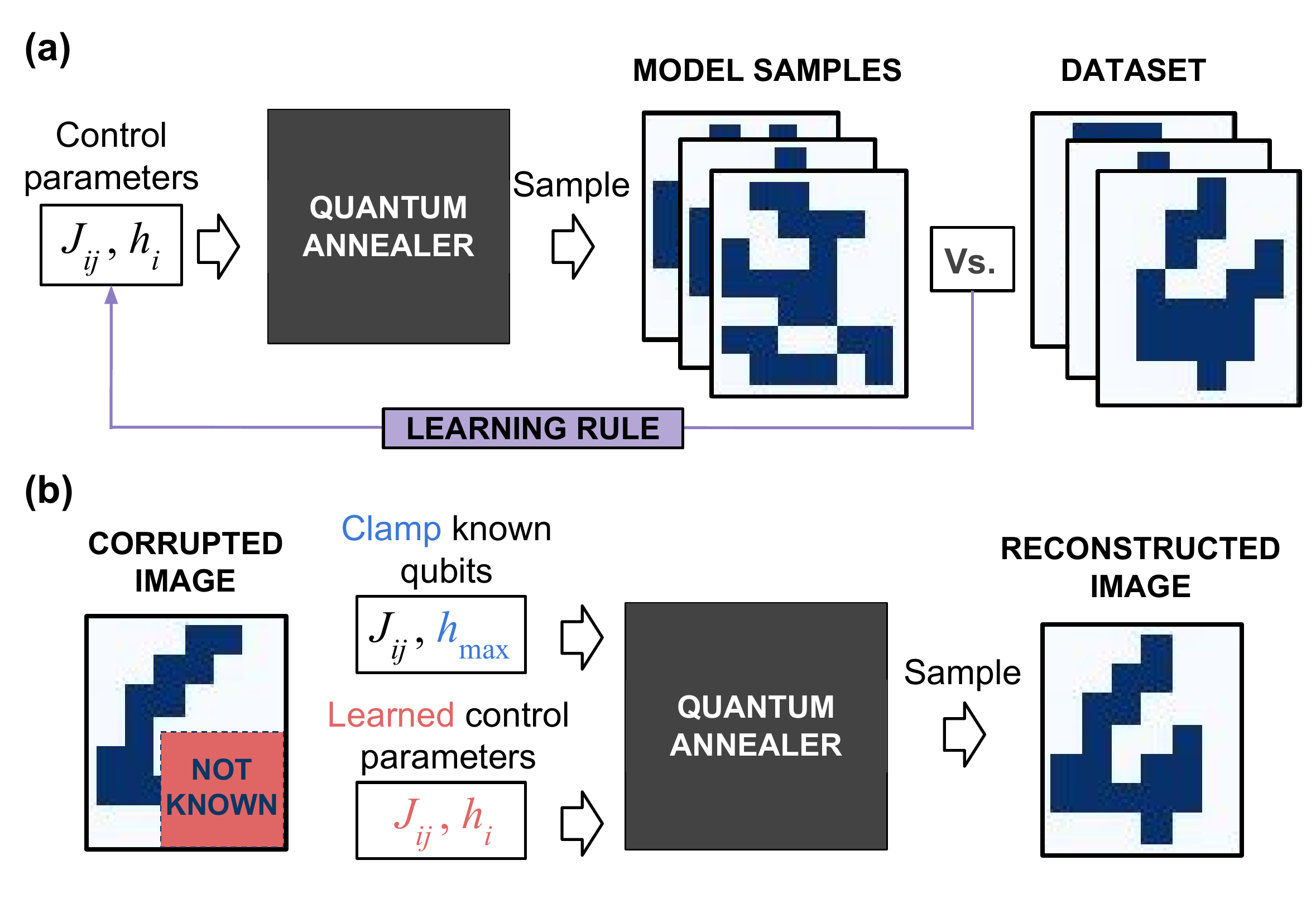}
\caption{{\it Quantum-assisted unsupervised learning.} (a) During the training phase, samples generated by the quantum annealer are compared with samples from the data set of, say, black-and-white images. The control parameters are then modified according to a learning rule (see Sec. III). This process is iterated a given number of times, also known as epochs. (b) After being trained we can use the quantum annealer, for instance, to reconstruct missing information in a data point, e.g., unknown values of some pixels (red region). To do this, we program the quantum annealer with the control parameters learned except for the fields of those qubits that represent known pixels. These fields are instead set to large values $h_{\rm max}$, so the qubits are clamped to the known values of the corresponding pixels. We then generate samples to infer the values of the unknown pixels.
}\label{f:design}
\end{figure}

Recently, there has been increasing interest in the potential that quantum computing technologies have for speeding up machine learning~\cite{neven2009nips, bian2010ising, Denil-2011,wiebe2012quantum, Pudenz-QIP-2013, Lloyd-arXiv-2013, Rebentrost-PRL-2014,wang2017quantum, 2015arXiv151203929Z,Lloyd-NatPhys-2014, schuld2016prediction, Wiebe-arXiv-2015, Benedetti-2016, Aaronson-2015,Adachi-arXiv-2015,chancellor2016maximum,Amin-arXiv-2016, kieferova2016tomography, kerenidis2016quantum, lamata2017basic, alvarez2016quantum, wittek2017quantum,Potok2017,Schuld-QML-2015, Romero2017, adcock2015advances, biamonte2016quantum,2017arXiv170708561C, PerdomoOrtiz2017, Benedetti2017b} or implementing more effective models~\cite{Gu2012}. This goes beyond the original focus of the quantum annealing computational paradigm~\cite{finnila1994quantum,kadowaki_quantum_1998,Farhi2001}, which was to solve discrete optimization problems~\cite{Gaitan2012,PerdomoOrtiz2012_LPF, Bian2014, OGorman-EPJST-2015,RieffelQIP2015,PerdomoOrtiz_EPJST2015,perdomo2015performance, Venturelli-JobShop-arXiv-2015, PerdomoOrtiz2017a}. Empirical results suggest that, under certain conditions, quantum annealing hardware samples from a Gibbs or a Boltzmann distribution~\cite{Benedetti-2016,Raymond-DWave-2016,Amin-arXiv-2015,Amin-arXiv-2016,Marc}. In principle, the user can adjust the control parameters so that the device implements the desired distribution. Figure ~\ref{f:design} shows an example of how, ideally, one could use a quantum annealer for the unsupervised task of learning handwritten digits. In practice, however, there exist device-dependent limitations that complicate this process. The most pressing ones are as follows~\cite{Benedetti-2016,Raymond-DWave-2016, bian2010ising, Denil-2011,Dumolin-2014}: (i) The effective temperature is parameter dependent and unknown, (ii) The interaction graph is sparse, (iii) the parameters are noisy, and (iv) the dynamic range of the parameters is finite. Suitable strategies to tackle all of these issues need to be developed before we can assess whether or not quantum annealers can indeed sample more efficiently than traditional techniques on conventional computers, or whether they can implement more effective models. A relatively simple technique for the estimation of parameter-dependent effective temperature was developed in Ref.~\cite{Benedetti-2016} and shown to perform well for training restricted Boltzmann machines. More recently, generalizations and alternative techniques have been introduced in Ref.~\cite{Raymond-DWave-2016}. In the context of machine learning, these techniques need to estimate temperature at each iteration, implying a computational overhead.

Here, we put forward an approach that completely sidesteps limitation (i), i.e., the need to estimate temperature at each iteration of the learning process. Furthermore, we propose a graphical model embedded in hardware that \textit{effectively} implements all pairwise interactions between logical variables representing the data set and that learns the parameter setting from data, improving on limitation (ii) . Since the essential components for estimating the gradient needed in the learning process take place on quantum hardware, our approach is more robust to the noise in the parameters, also improving on limitation (iii). 

Our work here is based on a quantum maximum-entropy model: a quantum Boltzmann machine with no hidden variables, whose learning in the classical limit is also known as the inverse Ising problem~\cite{Schneidman-Nature-2006,Ricci-Tersenghi-JSTAT-2012}. We show that the resulting models embedded in quantum hardware can model well both a coarse-grained binarized version of the optical recognition of handwritten digits (OptDigits)~\cite{Lichman:2013} and the synthetic bars-and-stripes (BAS) data set~\cite{MacKay-book-2002}.
Moreover, using data sets of configurations extracted from random instances of the Sherrington-Kirkpatrick model~\cite{Mezard-book-1987,Mezard-book-2009,Nishimori-book-2001}, we show that our model's generative performance improves with training and converges to the ideal value. These results provide strong evidence that quantum annealers can indeed be effectively used as samplers, and that their domain of application extends well beyond what was originally intended.

We emphasize that the objective of this work is not to address the question of quantum speedup in sampling applications but rather to provide the first clear experimental evidence that quantum annealers can be trained robustly and used in generative models for unsupervised machine-learning tasks. We use available techniques to transform the data set of interest into another data set with higher and redundant resolution, which is subsequently used to train models natively embedded in quantum hardware. We then use a gray-box model approach, which does not require us to estimate the effective temperature, nor the effective transverse field; this approach has the potential to correct for errors due to nonequilibrium deviations~\cite{Amin-arXiv-2015}, noise in the programmable parameters~\cite{PerdomoOrtiz_SciRep2016}, and sampling biases in available state-of-the-art devices~\cite{Mandra2017}. Hence, while the derivation of our quantum-assisted algorithm relies on the assumption that the quantum annealer is sampling from a Gibbs distribution, we do not expect that this assumption must be strictly valid for our algorithm to work well. Because we are optimizing a hard-to-evaluate convex function, the generative performance depends mostly on the quality and efficiency of the sampling required to estimate such function, an ideal situation for the purpose of benchmarking. Recently, our model and training methodology have been used to make progress in benchmarking quantum annealers for sampling~\cite{korenkevych2016benchmarking}, in contrast with the broadly explored topic of benchmarking combinatorial optimization.

The outline of this article is as follows. In Sec.~\ref{s:FC}, we describe how graphical models with {\it effectively} arbitrary pairwise connectivity can be embedded and realized in quantum hardware. Here, we emphasize the parameter-setting problem, which is essential for any implementation in hardware. In Sec.~\ref{s:LA}, we derive an algorithm that tackles the parameter-setting problem while learning the model from data. In Sec.~\ref{s:implementation}, we discuss the implementation details. In Sec.~\ref{s:results}, we describe the experiments performed on two synthetic data sets and a coarse-grained binarized version of the OptDigits data set; we show that the model introduced here, trained by using the D-Wave 2X (DW2X) hosted at NASA Ames Research Center, displays good generative properties and can reconstruct and classify data with good accuracy. In Sec.~\ref{s:conclusions}, we report the conclusions of our work, discuss the implementation of our approach in other hardware architectures such as the Lechner-Hauke-Zoller (LHZ) scheme~\cite{lechner2015quantum}, and present potential research directions.

\section{Hardware-embedded models} \label{s:FC}
\subsection{Quantum annealing and quantum models}
The dynamics of a quantum annealer are characterized by the time-dependent Hamiltonian 
\be\label{e:H}
\mathcal{H}(\tau) =   -A(\tau)\sum_{i\in\mathcal{V}}\hat{X}_i - B(\tau){\mathcal{H}}_P,
\ee
where $\tau=t/t_a$ is the ratio between time $t$ and annealing time $t_a$, while $A(\tau)$ and $B(\tau)$ are monotonic functions satisfying $A(0)\gg B(0)\approx 0$ and $B(1)\gg A(1)\approx 0$. The first term in Eq.~\eqref{e:H} above corresponds to the transverse field in the $x$ direction, characterized by the Pauli operators $\hat{ X}_i$ for each qubit $i$. The second term in Eq.~\eqref{e:H} corresponds to the problem-encoding Hamiltonian
\be\label{e:HP}
\mathcal{H}_P = -\sum_{(i,j)\in\mathcal{E}}  J_{ij} \hat{Z}_i \hat{Z}_j - \sum_{i\in\mathcal{V}} h_i \hat{Z}_i,
\ee
where $\hat{Z}_i$ refers to the $i$th qubit in the $z$ direction, which is defined on an interaction graph $\mathcal{G}=(\mathcal{V}, \mathcal{E})$. Here, $\mathcal{V}$ and $\mathcal{E}$ refer to the corresponding set of vertices and edges, respectively.

As discussed in Ref.~\cite{Amin-arXiv-2015}, the dynamics of a quantum annealer are expected to remain close to equilibrium until they slow down and start deviating away from equilibrium to finally freeze out. If the time between such dynamical slow-down and freeze-out is small enough, the final state of the quantum annealer is expected to be close to the quantum Gibbs distribution 	
\be
\label{e:rho}
\rho = \frac{e^{-\beta_{\rm QA} \mathcal{H}(\tau^\ast)}}{\mathcal{Z}},
\ee 
corresponding to the Hamiltonian in Eq.~\eqref{e:H} at a given point ${\tau=\tau^\ast}$, called freeze-out time. Here, ${\beta_{\rm QA}}$ is the physical temperature of the quantum annealer, and $\mathcal{Z}$ is the normalization constant. The density matrix in Eq.~\eqref{e:rho} is fully specified by the effective parameters ${W_{ij} = \beta\, J_{ij}}$, ${b_i = \beta\, h_i}$, and ${c = \beta\,\Gamma}$, where ${\beta= \beta_{\rm QA} B(\tau^\ast)}$ is the effective inverse temperature~\cite{Benedetti-2016, Amin-arXiv-2015} and ${\Gamma = A(\tau^\ast)/B(\tau^\ast)}$ is the effective transverse field.

If ${A(\tau^\ast)\ll B(\tau^\ast)}$, the final state of the quantum annealer is close to a {\em classical} Boltzmann distribution over a vector of binary variables ${\bz \in \{-1,+1\}^N}$,
\be\label{e:prob_dist}
P(\bz) = \frac{e^{-\beta E(\bz)}}{\mathcal{Z}},
\ee
where
\be\label{e:E}
E(\bz) = -\sum_{(i,j)\in\mathcal{E}}  J_{ij} z_i z_j - \sum_{i\in\mathcal{V}} h_i z_i
\ee
is the energy function given by the eigenvalues of $\mathcal{H}_P$ [see Eq.~\eqref{e:HP}].

The case where $A(\tau^\ast)$ cannot be neglected is less explored in the literature and allows the implementation of quantum Boltzmann machines~\cite{Amin-arXiv-2016}. All conditions described above, as well as the freeze-out time, depend on the specific instance of control parameters $J_{ij}$ and $h_i$ that are programmed. As shown in Sec.~\ref{s:LA}, our algorithm can also train hardware-embedded models despite these unknown dependencies. The potential to train quantum models~\cite{Amin-arXiv-2016, korenkevych2016benchmarking, kieferova2016tomography} opens new exciting opportunities for quantum annealing. These efforts resonate with foundational research interested in quantifying or identifying the particular computational resources that could be associated with quantum models~\cite{Gu2012, realpe2017quantum}.

\subsection{Enhancing modeling capacity}
In this section, we define the general setting to train a hardware-embedded probabilistic graphical model capable of representing graphs with arbitrary connectivity. Although we implement the general case of all-to-all connectivity as the most complex topology with pairwise interactions, working with models with simpler topologies can be easily represented with less numbers of qubits within this hardware-embedded setting. 

In combinatorial optimization, one seeks a configuration of binary variables $\bz$ associated with the lowest energy in Eq. \eqref{e:E}. The typical strategy to embed dense graphs in quantum hardware is to represent logical binary variables by subgraphs of the interaction graph of physical qubits. The value of  all control parameters should be fine-tuned such that the ground state of the original problem is preserved and therefore still favored in the physical implementation of the quantum annealing algorithm; this is known as the parameter-setting problem (see Sec.~\ref{ss:setting}-C and Refs.~\cite{choi2008minor,VentuPRX,perdomo2015performance}). 

In machine learning, the scenario is different. When learning a  model such as the one in Eq. \eqref{e:rho} or Eq. \eqref{e:prob_dist}, one seeks the configuration of control parameters $J_{ij}$ and $h_i$ that maximizes a suitable function of the data set. Notice that in combinatorial optimization problems, it is desirable to have the optimal configuration or ground state with probability close to one. In machine learning, however, all configurations are significant, as are their corresponding probabilities. By mapping the original problem to quantum hardware as routinely done in combinatorial optimization applications, we may end up implementing a distribution that differs from the intended one. 

An additional complication is that finding optimal parameters for a physical device is hampered by lack of precision, by noise, and by having to infer an instance-dependent effective temperature at each step of the learning. To avoid computing such an effective temperature at each learning iteration and to mitigate the effects of persistent biases~\cite{PerdomoOrtiz_SciRep2016}, lack of precision, and noise in the control parameters, we take a gray-box model approach. In other words, although we assume that samples generated by the quantum annealer are distributed according to Eq. \eqref{e:rho}, we do not need complete knowledge of the actual parameters being implemented. This leads to the condition that the first- and second-order moments of the model and data distributions should be equal for the parameters to be optimal. The resulting model is nevertheless tied to the specific machine being used. 

\begin{figure}
\includegraphics[width=.4\textwidth]{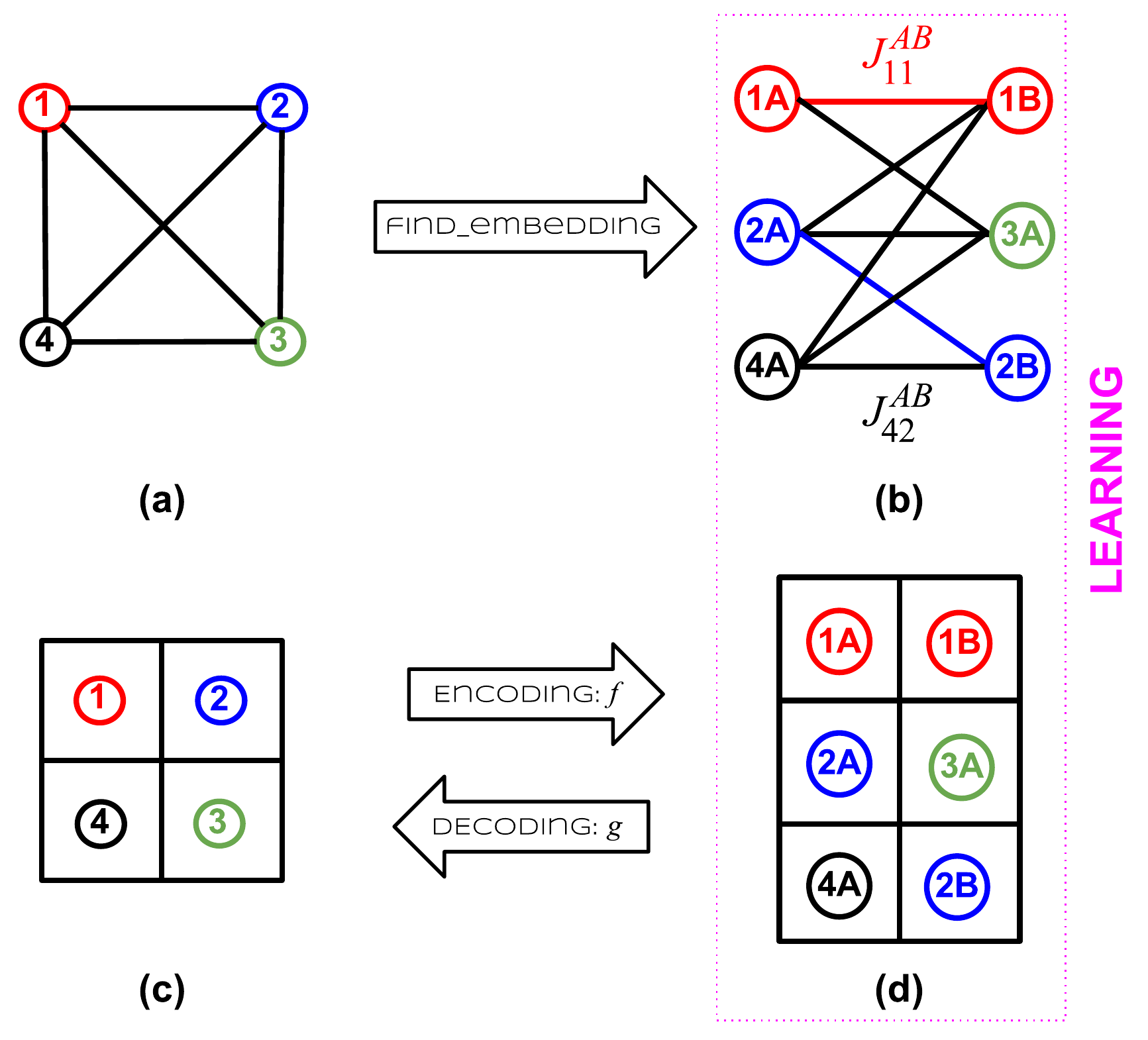}
\caption{{\it Hardware-embedded models.} (a) We first define a graph with arbitrary connectivity between the logical variables that directly encode the data set to be modeled; here, we show a fully connected graph on four variables as an example. Such a graph serves as a scaffolding to build hardware-embedded models with enhanced modeling capacity. (b) We then embed the scaffolding graph in quantum hardware by using minor embedding techniques; this requires the introduction of auxiliary qubits, couplings, and fields. In the example shown here, the logical variable $1$ (red) is encoded using two qubits, $1A$ and $1B$, which are connected by an auxiliary coupler $J_{11}^{AB}$; the same is true for variable $2$ (blue). The black links correspond to couplings between qubits representing different logical variables. In optimization problems, we are  given values for the couplings and fields on the graph of logical variables, as in diagram (a). To solve such optimization problems on a quantum annealer, we first have to pick values for all control parameters such that the ground state of the physical system coincides with the optimal solution of the problem being solved. The selection of control parameters can be done via handcrafted rules when information is available about the model~\cite{VentuPRX} or via heuristic approaches in the more general scenario~\cite{perdomo2015performance}. The optimal choice of control parameters, i.e., the one that maximizes the probability of finding the ground state, is known as the parameter-setting problem, and it is an open research question. In machine-learning applications, instead, the control parameters are not given but have to be found; they are the variables of the problem. In this case, embedding techniques are used both to transform the original data set [as shown in diagram (c)] into a data set with higher resolution due to redundant variables [as shown in diagram (d)] and to find a representation of such an extended dataset in hardware. This allows us to define encoding and decoding maps $f$ and $g$, respectively, to transform between the two data representations. Then, we forget about the scaffolding graph and train the hardware-embedded model on the extended data set. Thus, although the final hardware-embedded model might be interpreted as an embedding of a logical model, this is certainly not the case during learning, which automatically tackles the parameter-setting problem; in a sense, machine learning {\it is} parameter setting. While here we use standard embedding techniques to define the maps $f$ and $g$, such functions could, in principle, be learned from data, effectively automating the embedding problem, too. }\label{f:embedding}
\end{figure}

Using a generic logical graph as scaffolding, we associate each of its logical variables with a subgraph of the hardware interaction graph. This can be done by using existing minor embedding techniques; however, the parameter-setting problem remains. As an example, Figs.~\ref{f:embedding}(a) and \ref{f:embedding}(b) show a simple graph that cannot be directly implemented in DW2X hardware and a possible minor embedding, respectively. The additional couplings inside a subgraph are part of the hardware-embedded graphical model and have to be learned along with the model parameters that couple different subgraphs. In other words, the fine-tuning of all the couplings is done by the learning algorithm, which has the potential to learn corrections to the noise affecting the physical components, under the assumption that these defects still respect the direction of the gradient driving the learning algorithm. The embedding also allows us to map the data set into an extended data set with higher resolution, where some of the original variables are represented redundantly [see Figs.~\ref{f:embedding}(c) and ~\ref{f:embedding}(d)]. Then, the learning algorithm runs entirely on such extended space by training from scratch the whole hardware-embedded model on the extended data set. We now discuss, in more detail, the parameter-setting problem and how it is tackled in the type of machine-learning applications studied here.

\subsection{Parameter-setting problem}\label{ss:setting}
Let us define the parameter-setting problem as follows: Find values of control parameters to embed problems in hardware such that the performance of the device is ``optimal''. The meaning of optimal depends on the task of interest. In optimization problems, parameters are optimal if they provide the largest probability of finding a ground state~\cite{VentuPRX}. In the type of machine-learning applications considered here, parameters are optimal if samples generated by the device capture, as much as possible, the statistics of the data set. In a sense, machine learning {\it is} parameter setting. We discuss how this is quantified in Sec. \ref{s:LA}.  

Previous research~\cite{VentuPRX} suggests a possible mechanism underlying the parameter-setting problem. The authors investigated the Sherrington-Kirkpatrick model and found that the optimal choice of the additional parameters could be obtained by forcing both the spin glass and the ferromagnetic structures to cross the quantum critical point together during the annealing. Roughly speaking, the quantum phase transition happens when the energies associated with the problem-encoding system and the transverse field $\Gamma$ are of the same order of magnitude. This implies that the optimal embedding parameters are $O(J_{SG}\sqrt{N})$, where $N$ is the number of spins and $J_{SG}$ is the typical value of the couplings, that is, the standard deviation (see, e.g., Fig. 2 in Ref.~\cite{VentuPRX}). The intuition provided by this study does not necessarily apply to more realistic problems. In machine learning, even when starting from a fully connected model, the learning could still lead to a sparse final model.

As we discuss in Sec.~\ref{s:LA}, our approach lets the data guide the process by treating the whole quantum annealer as a neural network. In this case, both the inter- and intra-subgraph parameters are modified according to the statistical errors made by the quantum annealer in producing samples that resemble the data. This process implicitly corrects for noise and defects on the parameters, problems that are expected to affect any near-term quantum technology. The price to pay is a relatively small overhead as discussed in Sec.~\ref{s:LA}. 

In the following, we focus on hardware-embedded graphical models with {\it effective} all-to-all connectivity, as that is the most general case. Therefore, our derivations and the model proposed here include any topology with pairwise connectivity.\\

\subsection{Fully connected inspired models}\label{s:fcim}
The sparse interaction topology of state-of-the-art quantum annealers strongly limits their capacity to model complex data. For this reason, we use embedding strategies based on utilizing several qubits to represent a given variable in the data set. This amounts to transforming the data set of interest into a higher-resolution data set, with redundant variables, and modeling it with a hardware-embedded model.

More specifically, consider a binary data set ${\mathcal{D} = \{\bs^1, \dotsc , \bs^D\}}$, where each data point can be represented as an array of Ising variables, i.e., ${\bs^d = (s_1^d,\dotsc ,s_N^d)}$, with ${s_i^d\in\{-1,+1\}}$, for ${i=1,\dotsc , N}$. We refer to the $\bs$ variables as logical variables. We need to define a map $f$ from the data space to the qubit space that produces an extended binary data set ${\widetilde{\mathcal{D}} = \{\bz^1, \dotsc , \bz^D\}}$, where $\bz=f(\bs)$. In this work, we choose the map $f$ to replicate the state of each logical variable $s_i$ inside the corresponding subgraph $i$, i.e.,
\be \label{e:replica}
z_{i}^{(k)} = s_i, \quad \text{for} \quad k=1,\dots,Q_{i},
\ee
where $Q_{i}$ is the number of qubits in subgraph $i$. 

The task then turns into learning the parameters of a model on the extended data set $\widetilde{\mathcal{D}}$. To do this, we define a problem Hamiltonian over $M = \sum_{i=1}^{N} Q_{i}$ qubits,
\be \label{e:core}
\widetilde{\mathcal{H}}_P = -\frac{1}{2} \sum_{i,j=1}^N \sum_{k,l=1}^{Q_{i},Q_{j}} J_{ij}^{(kl)} \hat{Z}_{i}^{(k)} \hat{Z}_{j}^{(l)} - \sum_{i=1}^N \sum_{k=1}^{Q_{i}} h_{i}^{(k)} \hat{Z}_{i}^{(k)}.
\ee
Here, $N$ is the number of logical variables, which equals the number of subgraphs realized in hardware; $\hat{Z}_{i}^{(k)}$ is the Pauli matrix in the $z$ direction for qubit $k$ of subgraph $i$; $h_{i}^{(k)}$ is the local field for qubit $k$ of subgraph $i$; and $J_{ij}^{(kl)}$ is the coupling between qubit $k$ of subgraph $i$ and qubit $l$ of subgraph $j$. When $i=j$, it specifies the interactions within the subgraph, while when $i\neq j$, it specifies the interactions among subgraphs; $J_{ij}^{(kl)} =0 $ if there is no available interaction between the corresponding qubits in the quantum hardware. The binary variables $z_i^{(k)}$ encoding the extended data set can be interpreted as the eigenvalues of the Pauli matrix $\hat{Z}_{i}^{(k)}$.

After learning the parameters of $\widetilde{\mathcal{H}}_P$ in Eq.~\eqref{e:core}, we need a map $g$ from qubit space to data space that transforms samples generated by the quantum annealer into samples that resemble the original data set. Here, we choose $g$ to assign the state of the majority of physical variables in subgraph $i$ to the corresponding logical variable $s_i$, i.e.,
\be \label{e:majority}
s_{i} = \sign \left( \sum_{k=1}^{Q_{i}} z_i^{(k)} \right).
\ee
The rationale behind this choice is that, ideally, samples from the trained model are expected to have all qubits $z_i^{(k)}$ in a subgraph $i$ having exactly the same state, i.e., $z_i^{(k)} = z_i^{(l)}$ for $k,\, l = 1,\dotsc , Q_i$. In this case, we could pick whichever qubit $z_i^{(k)}$ as representative of the logical variable $s_i$, and this choice would be equivalent to the choice in Eq.~\eqref{e:majority}. However, we expect the choice in Eq.~\eqref{e:majority} to be more robust to the different sources of noise in quantum annealers by exploiting such a redundancy in the spirit of error-correction codes~\cite{Mezard-book-2009,Nishimori-book-2001}. While we have {\it a priori} fixed mappings $f$ and $g$ using embedding techniques, such functions could also be learned from data, as we will discuss elsewhere.

\section{Learning algorithm}\label{s:LA}
Let $\rho_\mcD$ be the diagonal density matrix whose diagonal elements encode the empiric data distribution. A quantum Boltzmann machine~\cite{Amin-arXiv-2016}, characterized by the density matrix $\rho$ defined in Eq.~\eqref{e:rho}, can be trained by minimizing the quantum relative entropy~\cite{kieferova2016tomography},
\be\label{e:SQ}
{S}\left(\rho_\mcD\middle\|\rho \right) = \mathrm{Tr}\rho_\mcD\ln \rho_\mcD - \mathrm{Tr}\rho_\mcD\ln \rho .
\ee
The learning rule is given by the equations
\BE
J_{ij}^{(kl)}(t+1) &=& J_{ij}^{(kl)}(t) + \eta \frac{\partial S}{\partial J_{ij}^{(kl)}}, \label{e:rule-J1}
\\
h_{i}^{(k)}(t+1) &=& h_{i}^{(k)}(t) + \eta\frac{\partial S}{\partial h_{i}^{(k)}},\label{e:rule-h}
\EE
where $t$ indicates the iteration and $\eta>0$ is the learning rate. Assuming we can neglect the dependence of the time lapsed between the dynamical slow-down and freeze-out in the instance of control parameters $J_{ij}^{(kl)}$ and $h_{i}^{(k)}$ programmed, we obtain
\BE
\frac{1}{\beta} \frac{\partial {S}}{\partial J_{ij}^{(kl)}} &=&  \bra \hat{Z}_{i}^{(k)} \hat{Z}_{j}^{(l)} \ket_{\rho_\mcD} - \bra \hat{Z}_{i}^{(k)} \hat{Z}_{j}^{(l)} \ket_{\rho},\label{e:Qgrad_J1}
\\
\frac{1}{\beta} \frac{\partial S}{\partial h_{i}^{(k)}} &=&  \bra \hat{Z}_{i}^{(k)} \ket_{\rho_\mcD} - \bra \hat{Z}_{i}^{(k)} \ket_{\rho} ,\label{e:Qgrad_h}
\EE
Here, $\bra\cdot \ket_{\rho_\mcD}$ denotes the ensemble average with respect to the density matrix $\rho_{\mcD}$ that involves {\it only} the data and is commonly referred to as the positive phase.  Similarly, $\bra\cdot\ket_\rho$ denotes the ensemble average with respect to the density matrix $\rho$ that exclusively involves the model and is called the negative phase.

If $A(\tau^\ast)\ll B(\tau^\ast)$ during our experiments, we are indeed dealing with classical models. Then, the learning rule above coincides with that for maximizing the average log-likelihood of the data~\cite{ackley1985learning}. 

However, the more general quantum case we just described provides a more accurate representation of the experiments we have performed, which are described below. To provide a strong argument as to which is the case, though, we need to carry out numerical simulations of the open quantum systems dynamics undergone by quantum annealers. We leave this for future work. However, our approach would also be valid for quantum annealers  capable of sampling from any desired fixed value of the transverse field, a capability that may be available in the near future.

The Hamiltonian in Eq.~\eqref{e:core} is designed to overcome connectivity limitations of hardware-embedded graphical models. In what follows, we show that the adaptation of standard learning procedures to the quantum maximum-entropy model proposed here works very well even in the presence of unknown hardware noise on couplings $J_{ij}^{(kl)}$ and local fields $h_{i}^{(k)}$. Moreover, we can learn suitable intra-subgraph couplings at a rate dictated by the contrast of the strength of pairwise correlations in the model and in the data, without the need for hard-coded values. 

Classically, the exact computation of the model's statistics is a computational bottleneck due to the intractability of computing the partition function and the exponential number of terms in the configuration space. An efficient approximation of the statistics is therefore required, and it is usually carried out by standard sampling techniques such as MCMC~\cite{ackley1985learning, Hinton-TechRep-2012}. In this work, we instead implement an algorithm that relies on the working assumption that quantum annealers can generate samples from a Gibbs-like distribution. However, even if this assumption is not strictly valid, our approach can still work as long as the estimated gradients have a positive projection in the direction of the true gradient. Quantum annealers have the potential to improve machine-learning algorithms in two ways: (i) by enabling the exploration of a wider class of models, i.e., quantum models, which some theoretical results~\cite{Gu2012} suggest may be able to capture higher complexity, and (ii) by speeding up the generation of samples. If the transverse field at the freezing point is negligible, the samples generated by the quantum annealer are expected to approximately follow a classical Boltzmann distribution. 

The learning procedure implemented by Eqs.~\eqref{e:Qgrad_J1} and \eqref{e:Qgrad_h} can be interpreted as quantum entropy maximization under constraints on the first- and second-order moments \cite{jaynes1957information,Jaynes-PhysRev-1957,Jaynes-book-2003}. In Ref.~\cite{chancellor2016maximum}, a maximum entropy approach was implemented on a D-Wave device in the context of information decoding, which is a hard optimization problem. Instead, we use quantum maximum-entropy inference for a hard machine-learning task, i.e., in unsupervised learning of generative models. 

Equation.~\eqref{e:rule-J1} implies that the intra-subgraph couplings $J_{ii}^{(kl)}$ increase at a varying rate proportional to ${1 -\small\langle \hat{Z}_i^{(k)}\hat{Z}_i^{(l)}\small\rangle}$, which, in principle, leads to infinite values in the long term. In practice, the rate of growth decreases as the learning progresses since the statistics of the samples generated by the quantum annealer resemble the data more and more. In general, the gradient-descent learning rule tends to produce too-large values for {\it all} the parameters because it pushes the model as much as possible towards a distribution with all the mass concentrated on the data. This problem in known as overfitting, and it is usually approached by regularization techniques. One regularization method may consist in penalizing large parameters by adding a term to Eq.~\eqref{e:SQ} accordingly. Another approach may be to employ a stopping criterion based on some measure of generalization or predictive capabilities of the model evaluated at each iteration on data not used during training. Under a proper choice of regularization, the intra-subgraph couplings utilized in our approach should not grow indefinitely anymore. However, regularization in the general setting of quantum machine learning is still an open research question.

Regarding the complexity of the algorithm, a fully connected model with $N$ logical variables has $O(N^2)$ parameters. When embedding such a fully connected model into a sparse graph like the Chimera graph of the DW2X, we end up with $O(N^2)$ qubits, but the number of parameters is still $O(N^2)$. This result occurs because we go from a dense graph of $N$ variables to a sparse graph of $O(N^2)$ variables. Each qubit in the DW2X interacts with, at most, six neighbors, so the number of additional parameters is a small constant factor. In our experiments, this factor is about 3 (see Table~\ref{t:embs}). Because of this factor, there is a small computational overhead for learning those intra-subgraph parameters. This overhead could be neglected because the main bottleneck is still in the generation of samples which is at least as hard as any non-deterministic polynomial time problem (NP-hard) An exact analog occurs in combinatorial optimization where a quadratic overhead is expected for embedding fully connected problems in hardware. In combinatorial optimization, such overheads are usually neglected because the main bottleneck is the NP-hard problem of reaching low-energy configurations.

A few additional remarks are in order: (i) The assumption that the model is based on a Gibbs distribution is reflected in that the second moment between two variables influences only the update of the corresponding coupling between them. If such a second moment increases (decreases), so does the corresponding coupling. This leaves open the possibility for the model to effectively self-correct for relatively small deviations from equilibrium, persistent biases, noise, and lack of precision, as long as the estimated gradient has a positive projection in the right direction, in the spirit of simultaneous perturbation stochastic approximation~\cite{Denil-2011,Spall-book-2003}.  (ii) The actual shape of a Gibbs distribution is instead characterized by the variables $\beta J_{ij}^{(kl)}$ and $\beta h_i^{(k)}$. Writing Eqs.~\eqref{e:rule-J1}~and~\eqref{e:rule-h} in terms of these new variables, we observe that the actual learning takes place at an effective learning rate that can vary since the effective temperature is instance dependent~\cite{Benedetti-2016}. (iii) The positive phases in Eqs.~\eqref{e:Qgrad_J1}~and~\eqref{e:Qgrad_h} are constants to be estimated exclusively from the data points, as there are no hidden units in our approach. In the case of generic models with hidden variables, this term becomes difficult to compute, in general, and we have to rely on approximations, e.g., via sampling or mean-field techniques. (iv) The related problem of estimating the parameters of a {\it classical} Ising model is called the inverse Ising problem~\cite{Schneidman-Nature-2006, Mezard-Mora-2009,Ricci-Tersenghi-JSTAT-2012}, and some of the main alternative techniques are mean-field and pseudo-likelihood methods.

\section{Implementation details}\label{s:implementation}
\subsection{Device and embeddings}
We run experiments on the DW2X quantum annealer located at NASA Ames Research Center. The device is equipped with $1152$ qubits interacting according to a graph known as Chimera connectivity. For the DW2X device hosted at NASA Ames, only $1097$ qubits are functional and available to be used. Assuming all $1152$ qubits were available, an efficient embedding schema ~\cite{choi2011minor} would allow us to implement a fully connected graph with up to $48$ logical variables. Since only $1097$ qubits are available, such a schema cannot be used, and the size of the largest fully connected model that can be implemented is reduced. For the embeddings of the instances studied here, we run the {\tt find\_embedding} heuristic~\cite{cai2014practical} offered by D-Wave's programming​ ​interface and use the best embedding found within the $500$ requested trials. We judge the quality of an embedding not only by the total number of physical qubits needed to represent the logical graph, but also by considering and preferring a smaller maximum subgraph size for the logical units. For example, in the case of the $46$-variable fully connected graph, we found an embedding with $917$ qubits and a maximum subgraph size of $34$. We selected, instead, an embedding with a larger number of qubits, $940$, but with a considerably smaller maximum subgraph size of $28$. (Figure~\ref{f:emb} in the Appendix shows the selected embedding, where each subgraph is represented by a number and a color.) Table~\ref{t:embs} shows details for each of the embeddings used in our experiments. Finally, the parameter range allowed by DW2X is $J^{(kl)}_{ij}\in[-1,+1]$ and $h^{(k)}_i\in[-2,+2]$. We initialized all the parameters to small values in $[ -10^{-6}, +10^{-6}]$ in order to break the symmetry.

\begin{table}[!htbp] \centering {
\resizebox{\columnwidth}{!}{%
\begin{tabular}{|c|c|c|c|c|c|c|}
    \hline
    \bfseries\thead{Logical\\variables} & \bfseries\thead{Physical\\variables} & \bfseries Min & \bfseries Max & \bfseries\thead{Chip\\usage} & \bfseries\thead{Logical\\parameters} & \bfseries\thead{Physical\\parameters}\\
    \hline
    15 & 76& 5  & 6 & 7\% & 120 & 252\\
    42 & 739 & 11  & 25 & 67\% & 903 & 2644 \\
    46 & 940 & 12  & 28 & 86\% & 1081 & 3389\\ 
    \hline
\end{tabular}} 
\caption{Main characteristics of the different embeddings used here, for each of the fully connected graphs. All embeddings were generated by the {\tt find\_embedding}~\cite{cai2014practical} heuristic provided by D-Wave's programming​ ​interface. The table includes information about the minimum (Min) and maximum (Max) subgraph size, the percentage of used qubits relative to those available (Chip usage), and the total number of parameters for the logical and physical graphs.}
\vspace*{-5mm}
\label{t:embs}
} \end{table}

\subsection{Data sets and preprocessing} \label{s:datasets}
We tested our ideas on the real OptDigits data set~\cite{Lichman:2013}, the synthetic BAS data set~\cite{MacKay-book-2002}, and a collection of synthetic data sets generated from random Ising instances.

\begin{figure}
\includegraphics[width=.45\textwidth]{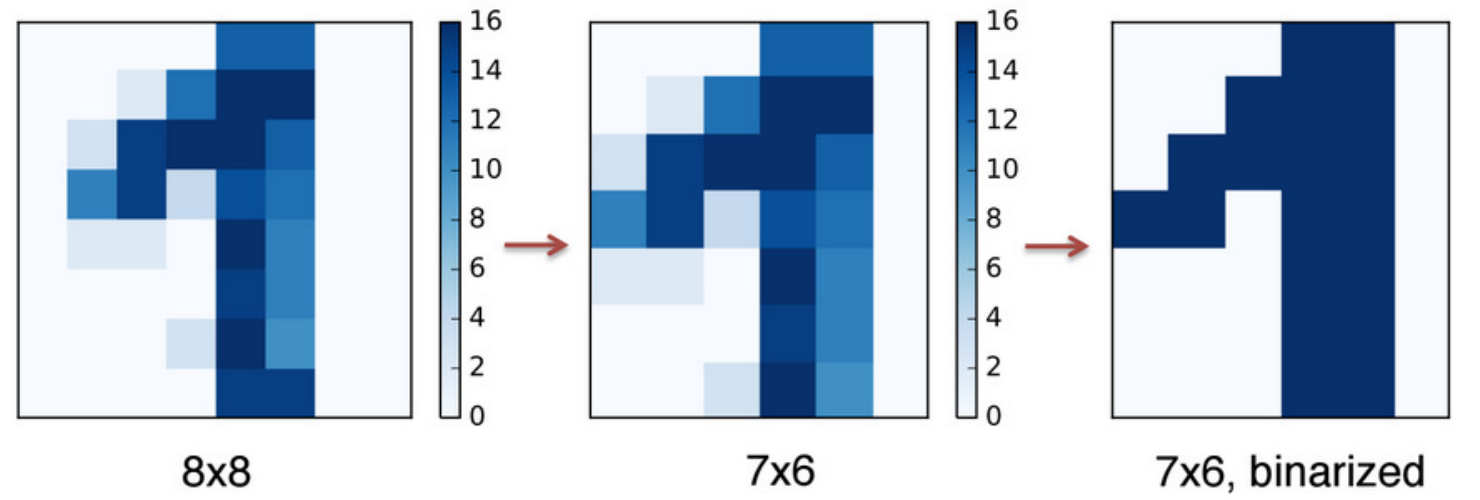}
\caption{{\it OptDigits preprocessing steps.} The original $8\times 8$ pictures are cropped to $7\times 6$ arrays by removing columns from the left and the right, as well as by deleting a row from the bottom. Finally, the four-bit gray scale is thresholded at the midpoint and binarized to $\{-1,+1\}$. Figure \ref{f:res2} (a) shows some pictures from the test set.   
}\label{f:prepro}
\end{figure}

The OptDigits data set requireds the preprocessing steps shown in Fig.~\ref{f:prepro}. First, each picture is $8\times 8$ and has a categorical variable indicating the class it belongs to. Using standard one-hot encoding for the class (i.e., $c^d_i=-1$ for $i\neq j$, $c^d_j=+1$, where $j$ indexes the class for picture $d$), we would need to embed a fully connected graph of $74$ variables, 64 for the pixels and 10 for the class, exceeding what we can embed in the DW2X. We removed the leftmost and rightmost columns as well as the bottom row from each picture, reducing the size to $7\times 6$ and retaining the readability. Second, we selected only four classes of pictures, those corresponding to digits ``one'' to ``four'', reducing the one-hot encoding to four variables. The four classes account for $1545$ pictures in the training set and $723$ pictures in the test set, and they are in almost equal proportion in both. Finally, the original four-bit gray scale of each pixel is thresholded at the midpoint and binarized to $\{-1,+1\}$ in order for the data to be represented by qubits in the DW2X. Figure \ref{f:res2} (a) shows some pictures from the test set.

The BAS data set consists of $N\times M$ pictures generated by setting the pixels of each row (or column) to either black ($-1$) or white ($+1$), at random. A reason to use this synthetic data set is that it can be adapted to the number of available variables in the DW2X. Having found an embedding for the $42$-variable fully connected graph, we generated a $7\times 6$ BAS data set consisting of $192$ pictures of $42$ binary variables each. Then, we randomly shuffled the pictures and split the data set into training and test sets of size $96$ each. Figure~\ref{f:res1}(a) shows some pictures from the test set. 

Finally, for the collection of synthetic data sets, we preferred to work with small-sized Ising instances that allowed us to carry out exhaustive computations. In particular, we chose $10$ random instances of a Sherrington-Kirkpatrick model with $N=15$ logical variables. Parameters $J_{ij}$ [cf. Eqs.~\eqref{e:prob_dist} and \eqref{e:E}] were sampled from a Gaussian with mean $\mu = 0$ and standard deviation $\sigma = \zeta / \sqrt{N}$, parameters $h_{i}$ were set to $0$, and the inverse temperature was set to $\beta=1$. In this setting, a spin-glass transition is expected when $\zeta_c = 1$ in the thermodynamic limit, although finite-size corrections are expected to be relevant for this small size. In order to obtain interesting structures within the probability distributions, we chose $\zeta = 2$ and verified that the overlap distribution~\cite{Mezard-book-1987,Mezard-book-2009} of each instance was indeed nontrivial. Moreover, we checked the performance of the closed-form solutions obtained by mean-field techniques in Ref. \cite{Ricci-Tersenghi-JSTAT-2012}. The mean-field method failed to produce (real-valued) solutions in seven out of the ten random instances, while it performed well in the remaining three instances, adding further evidence that these instances had nontrivial features in their energy landscape. Finally, we generated a training set of $D=150$ samples for each instance by exact sampling from its corresponding Boltzmann distribution. Table~\ref{t:data} summarizes the characteristics of each dataset used in our experiments.

\begin{table}[!htbp] \centering {
\begin{tabular}{|c|c|c|c|c|}
    \hline
    \bfseries Dataset & \bfseries Variables & \bfseries Training points & \bfseries Test points \\
    \hline
    OptDigits & $42+4$ & $1545$ & $723$ \\
    BAS $7\times6$* & $42$ & $96$ & $96$ \\
    Ising* & $15$ & $150$ & Not applicable \\
    \hline
\end{tabular}
\caption{Main characteristics of the datasets used here, i.e. number of variables, number of training points and number of test points when applicable. The * symbol indicates a synthetic dataset.}
\vspace*{-5mm}
\label{t:data}
} \end{table}

\subsection{Choice of hyperparameters}
We distinguish two kinds of hyperparameters: those associated with the device and those referring to the gradient. Device hyperparameters affect the time needed to obtain samples. We set them to their corresponding minimum values in order to obtain samples as fast as possible. Gradient hyperparameters come from advanced techniques known to improve generalization and speed up learning. We adopt standard $L2$ regularization for the pairwise interactions and momentum for all the parameters, hence introducing two hyperparameters in Eqs.~\eqref{e:rule-J1}~and~\eqref{e:rule-h} (see Ref. \cite{Hinton-TechRep-2012} for discussion about implementation details and best practices). For these hyperparameters, we tried a small grid of values and chose the value that would allow the quantum-assisted algorithm to produce visually appealing samples. All the experiments were performed using the hyperparameters shown in Table~\ref{t:hype}.

\begin{table}[!htbp] \centering {
\begin{tabular}{|c|c|c|}
    \hline
    \bfseries Domain & \bfseries Hyperparameter & \bfseries Value \\
    \hline
    \multirow{4}{*}{device} & annealing time & $5\mu s$ \\
    & programming thermalization & $1\mu s$ \\
    & readout thermalization & $1\mu s$ \\ 
    & auto scale & False \\ 
    \hline
    \multirow{3}{*}{gradient} & learning rate & $0.0025$ \\
    & $L2$ regularization & $10^{-5}$ \\
    & momentum & $0.5$ \\
    \hline
\end{tabular}
\caption{Settings used in all the experiments except those in Section \ref{ss:learn_ising}, where gradient hyperparameters were tuned.}
\vspace*{-5mm}
\label{t:hype}
}\end{table}

\section{Results}\label{s:results}
\subsection{Reconstruction of pictures}
The first task we address is verifying that the model is indeed able to learn the joint probability distribution of variables given a data set. One way to do this is to check whether the learned model can reconstruct corrupted pictures. To generate a reconstruction, we first need to enforce the value of each {\it correct} pixel to all qubits of the corresponding subgraphs, as illustrated in Fig.~\ref{f:design}(b). The qubits can be clamped to the desired value by using a strong local field in the corresponding direction. Notice that clamping variables in quantum annealers is somewhat different from its classical counterpart. Applying a strong local field to a qubit can substantially bias it towards a given value, but it still remains a dynamical variable. In classical computation, clamping a variable completely freezes it. We then generated samples from the learned model and assigned values to each corrupted pixel $s_i$ using the majority-vote map [Eq.~\eqref{e:majority}] for all qubits in the corresponding subgraph $i$. To further mitigate the noise associated with this, we generated multiple reconstructions, $100$ for each corrupted picture, and took a second majority vote over them. This approach is very robust as we did not observe any mismatch between the desired clamped variables and the corresponding readouts. We chose to interrupt the training of the models as soon as any of the parameters left the dynamic range of the device. Since the intra-subgraph couplings always increase, we expect these to be the first to get out of range, and we observed this result in the experiments described below. We use two data sets, OptDigits and BAS.

\subsubsection{Optical recognition of handwritten digits}
\begin{figure*}
\includegraphics[width=\textwidth]{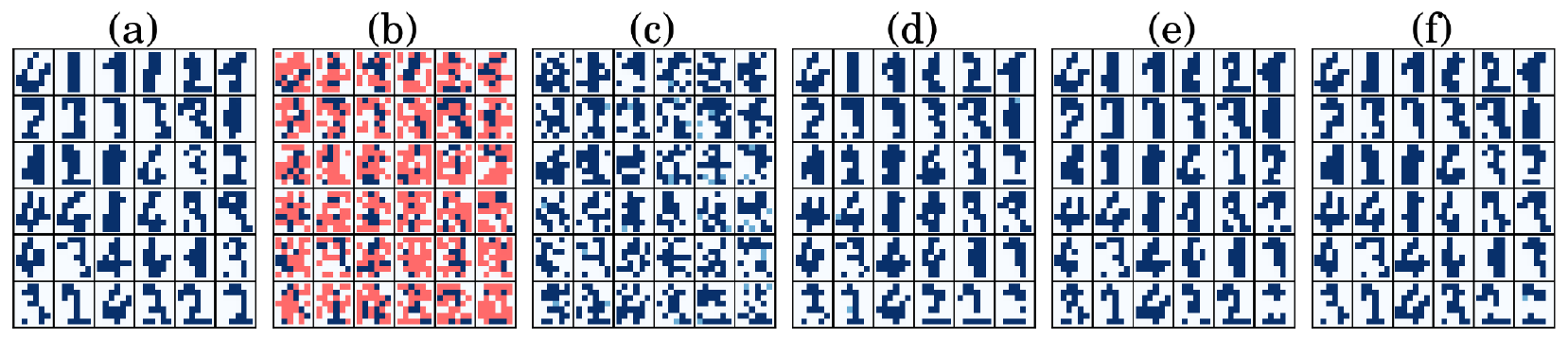}
\caption{{\it  OptDigits experiment.} (a) We show $36$ samples from the test set, with each pixel being either dark blue ($+1$) or white ($-1$). See Fig.~\ref{f:prepro} and the main text for a description of the preprocessing steps. (b) A uniform salt-and-pepper noise shown in red corrupts each picture. The model cannot use information from the red area. (c)-(f) Reconstructions obtained after $1$, $10$, $1000$, and $6000$ learning iterations. A light blue pixel indicates a tie of the majority vote over the corresponding subgraph. We can visually verify that the model has learned to generate digits. The learning stops at iteration $6000$ because further iterations would bring some parameters out of the dynamic range of the DW2X device.
}\label{f:res2}
\end{figure*}

We trained a model on the real data set OptDigits, a sample of which is shown in Fig. \ref{f:res2} (a). Since the training set contains a relatively large number of data points, we opted for a minibatch learning approach \cite{Hinton-TechRep-2012}, where $200$ data points were used at each iteration to compute the positive phase of the gradient. The negative phase is computed on $200$ samples from DW2X. We trained for $6000$ iterations, after which an intra-subgraph coupling went outside the dynamic range of the device. 

To evaluate the model, we added a $50$\% uniformly distributed ``salt-and-pepper'' noise [Fig.~\ref{f:res2}(b), red pixels] to each picture of the test set and used the model to reconstruct it. Notice that, given a corrupted picture, it is not always possible to obtain perfect reconstruction as multiple solutions could be correct. Therefore, we do not compute any error measure, but rather visually inspect the reconstructions. Figures~\ref{f:res2}(c)-~\ref{f:res2}(f) show some reconstructions obtained by models learned after $1$, $100$, $1000$, and $6000$ iterations, respectively. We can observe that qualitatively good reconstructions are already available from early stages of training. However, the large degree of corruption in the original image gives rise to things such as thicker reconstructions [Fig.~\ref{f:res2}(f), third row, fourth column], thinner reconstructions [Fig.~\ref{f:res2}(e), fourth row, second column], change of digit ``three'' to ``one'' [Fig.~\ref{f:res2}(e), third row, fifth column], among others. 

\subsubsection{Bars and stripes}
\begin{figure*}
\includegraphics[width=\textwidth]{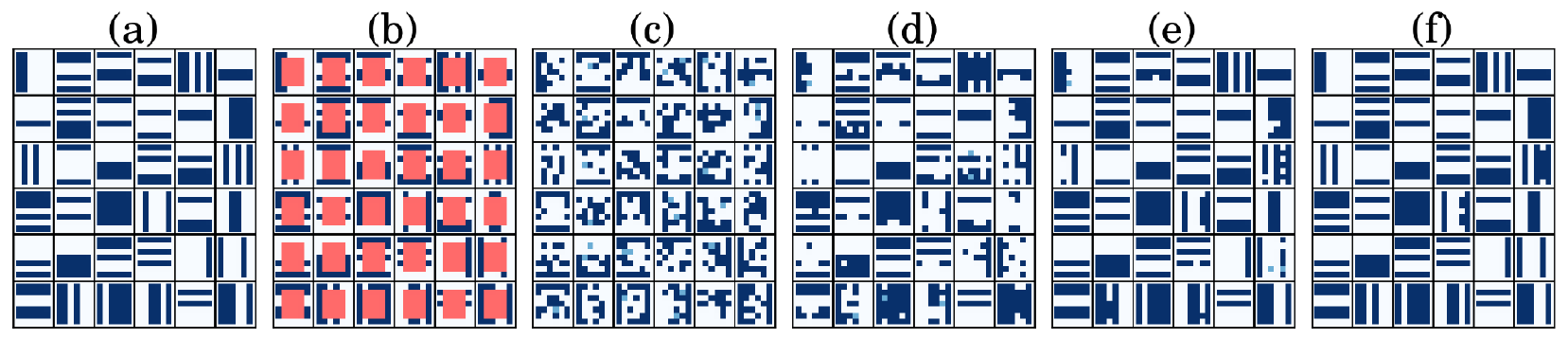}
\caption{{\it  BAS experiment.} (a) We show $36$ samples from the test set, with each pixel being either dark blue ($+1$) or white ($-1$). (b) A $5\times4$ block of noise shown in red corrupts each picture. The model cannot use information from the red area and yet the remaining pixels contain enough information to reconstruct the original picture. (c)-(f) Reconstructions obtained after $1$, $10$, $1000$, and $3850$ learning iterations. The average number of mistaken pixels is  $50$\% in (c), $18.6\%$ in (d), $2.95\%$ in (e), and finally $0.65\%$ in (f). This is an almost perfect reconstruction. The learning stops at iteration $3850$ because further iterations would bring some parameters out of the dynamical range of the DW2X device.  
}\label{f:res1}
\end{figure*}

We performed a similar test on the $7\times6$ BAS, a sample of which is shown in Fig. \ref{f:res1} (a). We computed the positive phase once using all $96$ training data points. Then, we ran the learning procedure, and for each iteration, we computed the negative phase out of $96$ samples obtained from the DW2X. The learning process stopped at iteration $3850$, after which an intra-subgraph coupling exceeded the maximum value allowed.

To evaluate the model, we blacked-out a $5\times4$ block [Fig.~\ref{f:res1}(b), red pixels corresponding to 47.6\% of the image] from each of the $96$ test pictures and used the model to reconstruct it. We can observe from Fig.~\ref{f:res1}(e) that reconstructed pictures are qualitatively similar to the original ones. To have a quantitative estimate of the quality of the reconstruction, we computed the expected number of incorrect pixel values (or mistakes) per reconstruction. After one iteration [Fig.~\ref{f:res1}(c)], we obtained a rate of $10.45$ mistakes out of $20$ corrupted pixels, corresponding to about $50$\% performance as expected. The number of mistakes decreased to $3.73$ ($18.6\%$) after $100$ iterations [Fig.~\ref{f:res1}(d)], $0.59$ ($2.95\%$) after $1000$ [Fig.~\ref{f:res1}(f)], and finally $0.13$ ($0.65\%$) at the end of training [Fig.~\ref{f:res1}(e)]. The latter result corresponds to almost perfect reconstruction. Notice​​ that,​ by​ definition, pictures​ from​​ the​​ test​ set​ are​ never​ used during​ training. Hence, these results provide evidence that the joint probability distribution of the pixels has been correctly modeled, and we can most likely rule out a simple memorization of the patterns.

\subsection{Generation and classification of pictures}
To investigate the generative and classification capabilities of the model, we introduced a one-hot encoding of the four classes of the OptDigits data set, therefore introducing four additional logical variables, for a total of $46$. We trained this larger model on the OptDigits data set, also including the classes.

We performed a simple classification task that does not require turning the generative model into a discriminative one by additional post-training. We classify each test picture as $c^{*} = \arg\max_{c} P(\bc|\bs)$, where $\bs$ is the vector encoding the pixels and $\bc$ is the vector encoding the classes. To approximate the probability, we clamped the subgraphs, by applying strong local fields, to the pixel values corresponding to the picture to be classified and sampled the four class variables from DW2X. We generated $100$ samples for each picture and assigned the picture to the most frequent class. After $6000$ learning iterations, this simple procedure led to an accuracy of $90$\% on the test pictures. This is a significant result, given that a random guess achieves $25$\% accuracy. However, it is to be expected that a fine-tuned discriminative model can achieve better accuracy.

\begin{figure*}
\includegraphics[width=.7\textwidth]{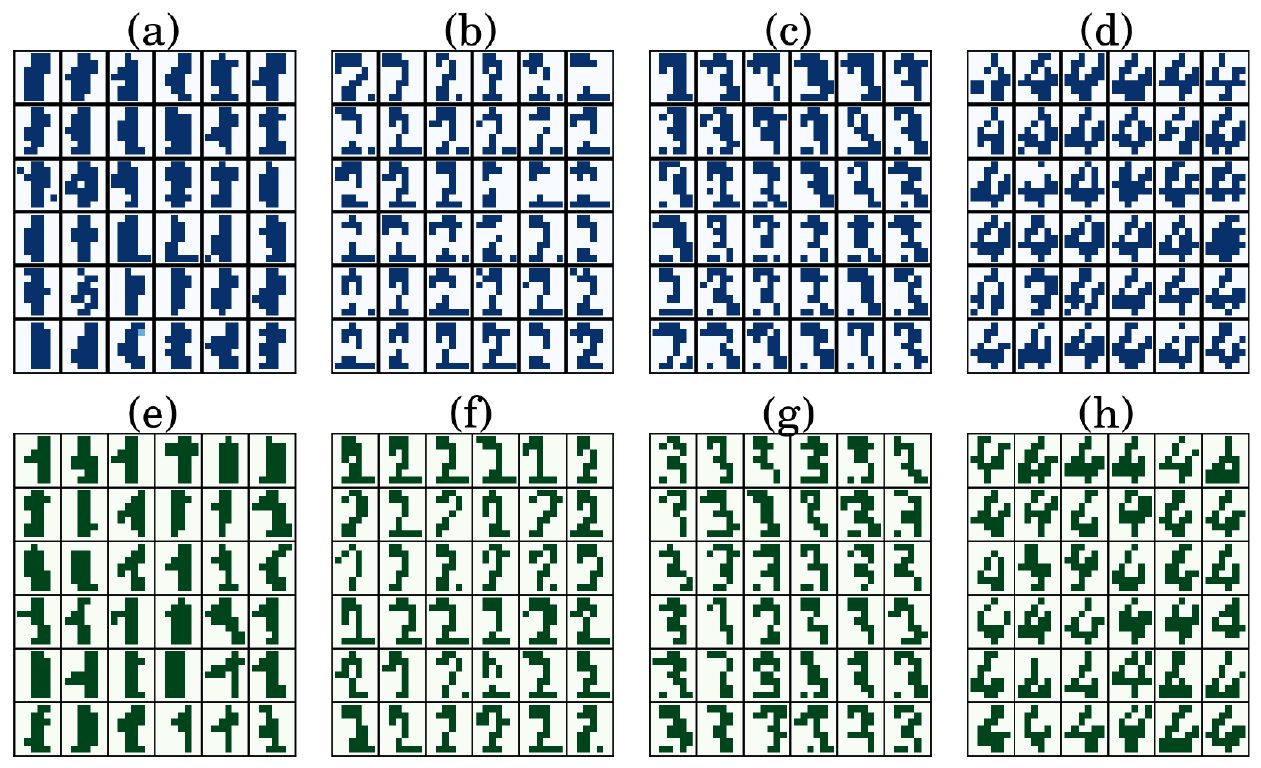}
\caption{{\it  Visual Turing test.} (a)-(h) The reader is invited to distinguish between human-generated pictures from the test set and machine-generated pictures sampled from the model. Columns identify classes one to four; rows identify the source--human or machine. The solution is given in Ref.~\cite{Note1}.
}\label{f:res3}
\end{figure*}

Finally, Fig.~\ref{f:res3} shows samples obtained from the DW2X by first setting the class variables, by applying strong local fields, to classes one to four (one class per column), along with human-generated pictures from the test set. Rows correspond to either human-generated pictures from the test set or machine-generated pictures. We defer the details of this visual Turing test to Ref.~\cite{Note1}. Machine-generated pictures are remarkably similar, though not identical, to those drawn by humans. Notice that human-generated digits may be ambiguous because of a variety of calligraphy styles encoded in low-resolution pictures. This ambiguity has been captured by the model, as shown by the machine-generated pictures.

\subsection{Learning of an Ising model}\label{ss:learn_ising}
In the previous section, we showed that quantum annealing can be used to successfully train the hardware-embedded models introduced here on both real and synthetic data sets of pictures. Here, we compare physical and logical models trained by quantum annealing (QA), simulated thermal annealing (SA)~\cite{kirkpatrick1983optimization}, and exact gradient. To simplify this task, we now deal with synthetic data sets composed of $D=150$ samples generated exhaustively from small-sized Boltzmann distributions as described in Sec.~\ref{s:datasets}. This is similar in spirit to the approach usually taken in the literature on the inverse Ising problem~\cite{Schneidman-Nature-2006,Ricci-Tersenghi-JSTAT-2012}. However, we do not quantify the quality of the trained model by the quadratic error between the parameters of the original model and those obtained by the learning algorithms, as it is usually done, for three reasons: (i) The physical model implemented in quantum hardware has a larger number of parameters than the logical model from which the data are generated, and a direct comparison is not straightforward. (ii) In our gray-box model approach, we do not have direct access to the effective parameters implemented in the quantum annealer, so we have to estimate the effective temperature that can introduce errors. (iii) To our knowledge, there is no direct connection between generic distances in parameter space, as measured by the quadratic error, and distances in probability space, which are those that have actual operational meaning, except perhaps when the parameters are close enough. Indeed, to measure distances in parameter space that correspond to distances in probability space it is necessary to use the Fisher information metric. For instance, it is known that, close to a critical point, a slight variation in the parameters can lead to drastically different probability distributions~\cite{Mastromatteo-JSTAT-2011}. 

Instead, our evaluation strategy exploits the fact that we have full knowledge of the probability distribution $Q(\bs)$ that generated the data. At each learning iteration, or epoch, we sample a set $\mathcal{S}=\{\bs^{(1)},\dotsc , \bs^{(L)}\}$ of $L$ points from the model $P(\bs)$ and evaluate the average log-likelihood that such samples were generated by $Q(\bs)$,
\be
\label{e:test}
\begin{split}
\Lambda_{\rm av}(\mathcal{S})&=\frac{1}{L}\sum_{\ell = 1}^L\log Q( \bs^{(\ell)}) \\
&= -\beta\frac{1}{L}\sum_{\ell = 1}^L E(\bs^{(\ell)}) - \log Z(\beta) ;
\end{split}
\ee
for simplicity, we chose $L=D=150$. Notice that Eq.~\eqref{e:test} requires full knowledge of the distribution that generated the data. This is unfeasible for real data sets since the whole point of learning a model is precisely that we do not know the true underlying distribution. However, this proxy is related to the generalization properties of the trained model since it corresponds to the likelihood that new arbitrary samples generated by the model were actually generated by the true underlying distribution. We expect this to be a faithful proxy since achieving good generalization performance is the main objective of machine-learning techniques. However, we should take into account that $\Lambda_{\rm av}(\mathcal{S})$ is not expected to be maximized by the generated samples but rather to match the value $\Lambda_{\rm av}(\mcD)$ of the original data set.

\begin{figure*}
\includegraphics[width=.85\textwidth]{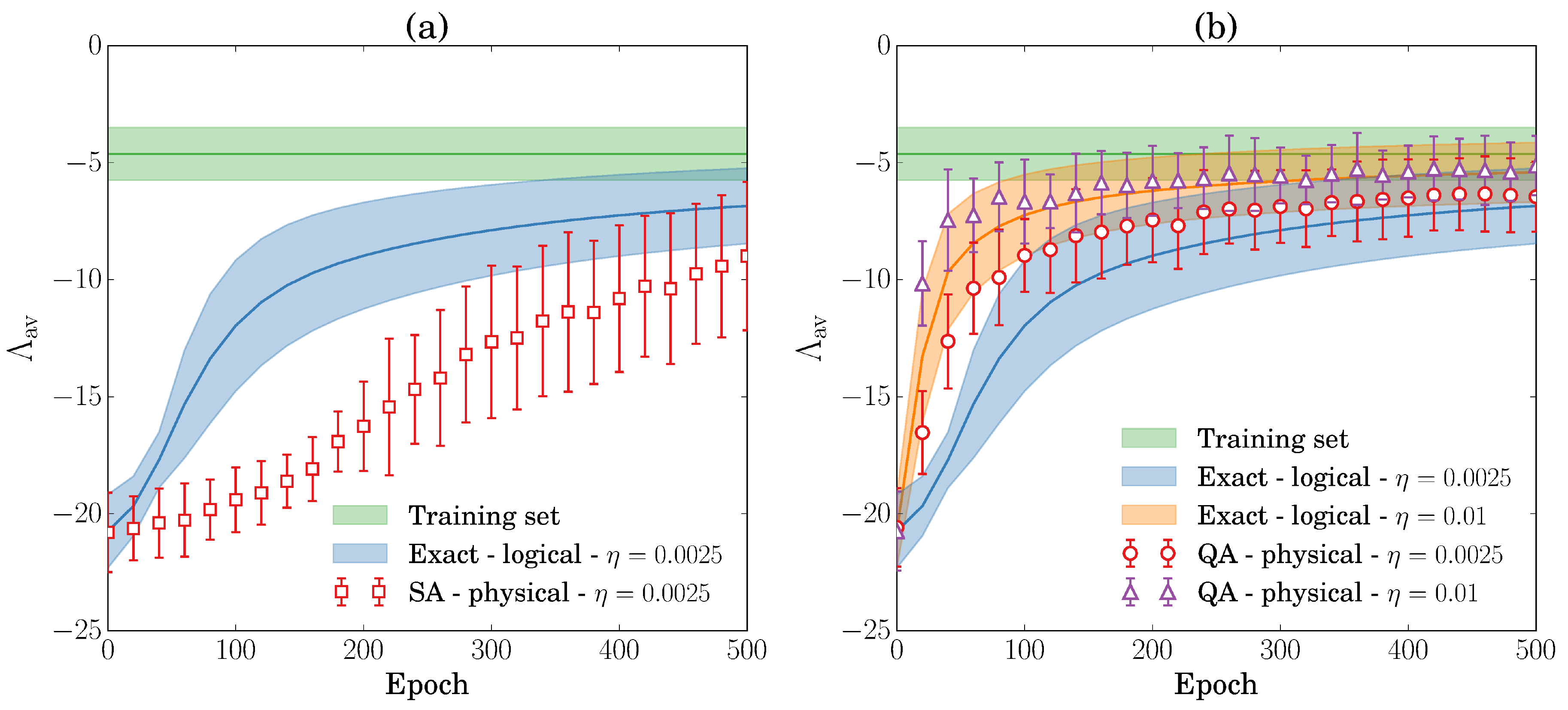}
\caption{\textit{Comparison of different learning settings.} The plots show mean and $1$ standard deviation of the proxy $\Lambda_{\rm av}$ for $10$ random instances and for different learning procedures. We use exact gradient for the $15$-variable logical graph and quantum annealing (QA) or simulated thermal annealing (SA) for the corresponding $76$-qubit physical graph. A learning procedure is considered successful if it can generalize, that is, if it matches the proxy of the training set (green band). (a) The logical model (blue band) matches faster than the physical model (red squares) when the same learning rate is used. This suggests that the larger number of parameters does not help the physical model. (b) Quantum annealing on the physical graph (red circles) enables faster matching than exact gradient on the logical graph (blue band) when the same learning rate $\eta = 0.0025$ is used. However, the exact-gradient procedure equipped with a larger learning rate $\eta = 0.01$ (orange band) outperforms the quantum-assisted algorithm. In turn, the quantum-assisted algorithm outperforms all other learning procedures when  equipped with the larger learning rate $\eta = 0.01$ (purple triangles).  Notice that neither the computation of $\Lambda_{\rm{av}}$ nor the exact-gradient learning is tractable, in general.}\label{f:res5}
\end{figure*}

In this set of experiments, we performed $500$ learning iterations and did not use gradient enhancements such as momentum and regularization in order to simplify the quantitative analysis. 
First, we verified whether the larger number of parameters in the physical graph provides a practical advantage against the logical models. While exact gradient calculations are feasible in the $15$-variable logical graph, they are infeasible for the $76$-qubit physical graph considered here (see details in Table~\ref{t:embs}). We opted for a sampling procedure based on SA where each sample follows its own independent linear schedule, therefore avoiding the problem of autocorrelation among samples. We used a linear schedule ${\beta(t)=t/t_{max}}$ for the inverse temperature and performed a preliminary study in order to set the optimal number of Monte Carlo spin flips per sample, $t_{max}$. We incrementally increased this number and observed the change in learning performance via the proxy $\Lambda_{\rm av}$. We choose ${t_{max}=15200}$ Monte Carlo spin flips, as multiples of this number did not result in improved learning speed nor in better values of $\Lambda_{\rm av}$. We expect this procedure to be essentially equivalent to exact gradient within the $500$ learning iteration considered here. Figure~\ref{f:res5} shows mean and $1$ standard deviation of the performance indicator for the $10$ synthetic instances considered here. Figure~\ref{f:res5}(a) indicates that SA-based learning on the physical graph (red squares) is slower than exact gradient learning on the logical graph (blue band) when the same learning rate is used. Even though both methods approach the optimal $\Lambda_{\rm av}$ of the data set (green band), the larger number of parameters does not speed up learning. Despite this, Fig.~\ref{f:res5}(b) shows that quantum-assisted learning with $\eta = 0.0025$ (red circles) outperforms exact gradient. This indicates that a varying effective learning rate could be induced by the instance-dependent effective temperature~\cite{Benedetti-2016} at which a quantum annealer samples. Indeed, by increasing the learning rate of the exact gradient method to $\eta=0.01$ (orange band), we were able to outperform quantum-assisted learning. In turn, however, quantum-assisted learning can outperform exact gradient if the same larger learning rate is used (purple triangles). The fast initial learning could also be caused by a nonvanishing transverse field at the freeze-out point (see Sec.~\ref{s:LA} above for a discussion). 
Because of the interplay between effective temperature and learning rate, the experiments presented here cannot confirm nor rule out the presence of these quantum effects. Open-quantum-systems simulations on small systems might give us greater control and allow us to have further insights into the interplay of the mechanisms proposed here. We leave this task for future work.

\section{Conclusions}\label{s:conclusions}
Whether quantum annealing can improve algorithms that rely on sampling from complex high-dimensional probability distributions, or whether they can provide more effective models are important open research questions. However, quantum annealers face several challenges that need to be overcome before we can address such a question from an experimental perspective. Besides the problem of proper temperature estimation, which has been addressed recently~\cite{Benedetti-2016,Raymond-DWave-2016}, some of the most pressing challenges are sparse connectivity which limits the capacity of the models that can be implemented, low precision, and limited range of the control parameters, as well as different sources of noise that affect the performance of state-of-the-art quantum annealers~\cite{Dumolin-2014}. 

By combining standard embedding techniques with the data-driven automatic setting of embedding parameters, we substantially improve the robustness and the complexity of machine-learning models that can be modeled with quantum annealers. By working on a gray-box model framework, which requires only partial information about the actual distribution from which the quantum annealer is sampling, this approach also avoids the need for estimating temperature during the learning of the models and has the potential to help mitigate the different sources of noise on the device. The resulting model can be interpreted as a visible-only quantum Boltzmann machine with all pairwise interactions among logical variables. We validated our ideas qualitatively by training the fully connected hardware-embedded model for reconstruction and generation of pictures, and quantitatively by computing a proxy on data sets extracted from randomly generated Boltzmann distributions.

Another advantage of our approach is that the learning rules are embedding-agnostic. More precisely, the underlying hardware embedding for the scaffolding logical model can be found by either heuristic algorithms~\cite{cai2014practical} or by known efficient schemes~\cite{choi2011minor,Klymko2012}, and the learning strategy is the same. While we have {\it a priori} fixed mappings $f$ and $g$ using embedding techniques, such functions could also be learned from data, as we will discuss elsewhere.

Furthermore, the strategy for training can be straightforwardly extended to other proposed hardware architectures, such as the LHZ scheme~\cite{lechner2015quantum}. More specifically, the data from the machine-learning task can be easily mapped to the physical qubits of that scheme by following the equivalent of our Eq.~\eqref{e:replica}. One difference is that the gradient updates [see Eqs.~\eqref{e:Qgrad_J1} and \eqref{e:Qgrad_h}] for the programmable parameters in this case will involve the updates of bias terms and the penalties for the quartic terms, under that choice of hardware implementation. This does not pose any challenges with our approach either, and the final results of the same iterative learning procedure detailed here would be a trained quantum or classical model. By using this gray-box model, one can also get samples from a LHZ-type device and use it for useful tasks such as the digit reconstruction or generation as illustrated in this work. The question of whether there is any advantage of either implementation for the machine-learning tasks proposed here is a question that would need to be addressed in future work.  

Natural extensions of the model will be inclusion of latent variables, also known as hidden units, support for continuous variables, and the development of techniques for the quantum annealer to also learn the embedding from data. Hidden units are needed, for example, if visible patterns require constraints that cannot be enforced by pairwise interactions alone \cite{ackley1985learning}. Continuous variables are needed for a correct modeling of real data sets. This has been the focus of recent work~\cite{Benedetti2017b,PerdomoOrtiz2017}, where we used the same gray-box model developed here but on a fully connected graph of $60$ hidden units (instead of visible units). We performed experiments on a hardware-embedded model with $1644$ qubits, further supporting the robustness-to-noise claims in this work and the value of this approach as a template for other quantum-assisted frameworks.

Another possible direction for future work is the extension of our learning algorithm to more general, possibly nonequilibrium, distributions. As we discussed, our learning algorithm might still work when there are relatively small deviations from the thermal distribution we assumed, as long as the estimated gradient has a positive projection on the direction of the true gradient. Indeed, if we had no information on the state reached by the quantum annealer, we would have to rely on model-free (e.g., black-box) techniques based, for instance, on randomly choosing a direction to update the control parameters, which may be highly inefficient~\cite{Denil-2011,Spall-book-2003}. On the other hand, if we had complete knowledge of the final state reached by the quantum annealer, we could benefit from model-based techniques, as we have done here. A possible hybrid algorithm may be based on a model, e.g., a Gibbs distribution, that captures the most relevant features of the quantum annealer state and some unknown corrections. In this way, the algorithm may choose a direction in parameter space informed by the model, instead of just randomly, and use the black-box techniques to correct for mistakes.

From a more fundamental perspective, several key questions remain open: When and why could the quantum annealer do better than classical MCMC approaches, or when and why could it provide more effective models? Our results show that the quantum-assisted learning algorithm has a faster learning during the initial stage, in the scenario where both classical (exact gradient estimation and SA) and our hybrid quantum-classical approach are set under the same conditions in terms of hyperparameters. Given that an instance-dependent effective temperature can imply a varying learning rate, this faster learning is probably due to the quantum-assisted algorithm automatically adjusting its learning rate. In this respect, it is important to investigate if such a learning schedule is optimal and, if so, whether it can be effectively simulated by classical means. Still, we cannot discard that some nontrivial quantum effects play a role here. Indeed, as pointed out in Ref.~\cite{kieferova2016tomography}, and as we further discussed above, the learning rules for classical and quantum Boltzmann machines coincide when there are no hidden variables. The potential to train quantum models~\cite{Amin-arXiv-2016, korenkevych2016benchmarking, kieferova2016tomography} opens new exciting opportunities for quantum annealing. These efforts resonate with foundational research interested in quantifying or identifying the computational resources that could be associated with quantum models~\cite{Gu2012, realpe2017quantum}. 

Arguably, this question of whether or not there is quantum speedup in sampling applications is one of the most important questions propelling our research. Years of experience accumulated with the use of quantum annealers for combinatorial optimization suggest that the answer may not be straightforward~\cite{Job2017,Katzgraber2017}, with the first comprehensive benchmarking study on an industrial application performed only recently~\cite{PerdomoOrtiz2017a}. Benchmarking quantum annealing for machine learning can be approached by following well-established guidelines used in optimization (see Ref.~\cite{ronnow2014defining}). However, the iterative nature of most machine-learning applications makes the task far more time-consuming. Almost all the hyperparameters (e.g., learning rate, annealing time, number of samples per iteration, etc.) can be adjusted throughout the learning, hence requiring us to find an optimal schedule for both classical and quantum algorithms. To obtain acceptable statistics, the study should be carried out on several data sets and different system sizes, where the time required to optimize the hyperparameters above grows quickly with the system size. In nonconvex problems (in parameter space), even if the samples used at each iteration are of high quality, the learning algorithm can find suboptimal solutions. In convex problems like the one we considered here, the performance of the learning algorithm mostly relies on the quality of the samples. This makes our approach appealing for the purpose of benchmarking. Still it is required to assess the quality of the whole distribution of states and not just the ground state as in combinatorial optimization applications. In this work, we focus on providing a proof-of-principle demonstration and experimental evidence that quantum annealers can be used for complex machine-learning tasks, such as in the case of unsupervised generative modeling on fully visible, probabilistic, graphical models with arbitrary pairwise connectivity. We hope this work continues opening new opportunities for quantum annealing and, more broadly, for quantum machine-learning research. 

\section*{Acknowledgements}
This work was supported in part by the AFRL Information Directorate under Grant No. F4HBKC4162G001, the Office of the Director of National Intelligence (ODNI), and the Intelligence Advanced Research Projects Activity (IARPA), via IAA 145483. The views and conclusions contained herein are those of the authors and should not be interpreted as necessarily representing the official policies or endorsements, either expressed or implied, of ODNI, IARPA, AFRL, or the U.S. Government. The U.S. Government is authorized to reproduce and distribute reprints for Governmental purpose notwithstanding any copyright annotation thereon. M. B. was partially supported by the UK Engineering and Physical Sciences Research Council (EPSRC) and by Cambridge Quantum Computing Limited (CQCL).

\onecolumngrid
\appendix

\section{Example}
\begin{figure}[!h]
\includegraphics[width=0.62 \columnwidth]{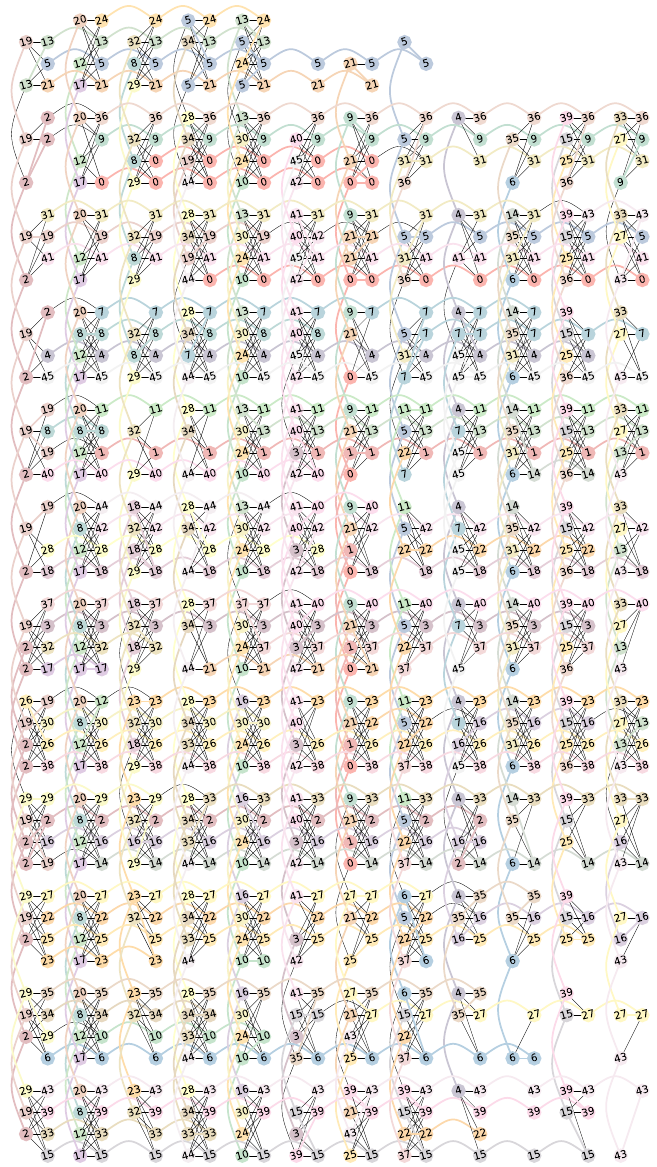}
\caption{{\it Embedding.} We show $46$ logical variables embedded into DW2X's chimera graph using $940$ physical variables. Qubits belonging to a logical variable are identified by the same number and linked by edges of the same color. This embedding uses $86\%$ of DW2X's qubits.}\label{f:emb}
\end{figure}

Figure~\ref{f:emb} shows the embedding of a fully connected graph with $46$ logical units into $940$ physical qubits.

\newpage
\twocolumngrid

%


\begin{thebibliography}{85}%
\makeatletter
\providecommand \@ifxundefined [1]{%
 \@ifx{#1\undefined}
}%
\providecommand \@ifnum [1]{%
 \ifnum #1\expandafter \@firstoftwo
 \else \expandafter \@secondoftwo
 \fi
}%
\providecommand \@ifx [1]{%
 \ifx #1\expandafter \@firstoftwo
 \else \expandafter \@secondoftwo
 \fi
}%
\providecommand \natexlab [1]{#1}%
\providecommand \enquote  [1]{``#1''}%
\providecommand \bibnamefont  [1]{#1}%
\providecommand \bibfnamefont [1]{#1}%
\providecommand \citenamefont [1]{#1}%
\providecommand \href@noop [0]{\@secondoftwo}%
\providecommand \href [0]{\begingroup \@sanitize@url \@href}%
\providecommand \@href[1]{\@@startlink{#1}\@@href}%
\providecommand \@@href[1]{\endgroup#1\@@endlink}%
\providecommand \@sanitize@url [0]{\catcode `\\12\catcode `\$12\catcode
  `\&12\catcode `\#12\catcode `\^12\catcode `\_12\catcode `\%12\relax}%
\providecommand \@@startlink[1]{}%
\providecommand \@@endlink[0]{}%
\providecommand \url  [0]{\begingroup\@sanitize@url \@url }%
\providecommand \@url [1]{\endgroup\@href {#1}{\urlprefix }}%
\providecommand \urlprefix  [0]{URL }%
\providecommand \Eprint [0]{\href }%
\providecommand \doibase [0]{http://dx.doi.org/}%
\providecommand \selectlanguage [0]{\@gobble}%
\providecommand \bibinfo  [0]{\@secondoftwo}%
\providecommand \bibfield  [0]{\@secondoftwo}%
\providecommand \translation [1]{[#1]}%
\providecommand \BibitemOpen [0]{}%
\providecommand \bibitemStop [0]{}%
\providecommand \bibitemNoStop [0]{.\EOS\space}%
\providecommand \EOS [0]{\spacefactor3000\relax}%
\providecommand \BibitemShut  [1]{\csname bibitem#1\endcsname}%
\let\auto@bib@innerbib\@empty
\bibitem [{\citenamefont {LeCun}\ \emph {et~al.}(2015)\citenamefont {LeCun},
  \citenamefont {Bengio},\ and\ \citenamefont {Hinton}}]{LeCun-Nature-2015}%
  \BibitemOpen
  \bibfield  {author} {\bibinfo {author} {\bibfnamefont {Yann}\ \bibnamefont
  {LeCun}}, \bibinfo {author} {\bibfnamefont {Yoshua}\ \bibnamefont {Bengio}},
  \ and\ \bibinfo {author} {\bibfnamefont {Geoffrey}\ \bibnamefont {Hinton}},\
  }\bibfield  {title} {\enquote {\bibinfo {title} {{Deep learning}},}\
  }\href@noop {} {\bibfield  {journal} {\bibinfo  {journal} {Nature}\ }\textbf
  {\bibinfo {volume} {521}},\ \bibinfo {pages} {436 -- 444} (\bibinfo {year}
  {2015})}\BibitemShut {NoStop}%
\bibitem [{\citenamefont {Ghahramani}(2015)}]{Zoubin-Nature-2015}%
  \BibitemOpen
  \bibfield  {author} {\bibinfo {author} {\bibfnamefont {Zoubin}\ \bibnamefont
  {Ghahramani}},\ }\bibfield  {title} {\enquote {\bibinfo {title}
  {{Probabilistic machine learning and artificial intelligence}},}\ }\href@noop
  {} {\bibfield  {journal} {\bibinfo  {journal} {Nature}\ }\textbf {\bibinfo
  {volume} {521}},\ \bibinfo {pages} {452 -- 459} (\bibinfo {year}
  {2015})}\BibitemShut {NoStop}%
\bibitem [{\citenamefont {Bengio}\ and\ \citenamefont
  {Courville}(2016)}]{Bengio-Book}%
  \BibitemOpen
  \bibfield  {author} {\bibinfo {author} {\bibfnamefont {Ian
  Goodfellow~Yoshua}\ \bibnamefont {Bengio}}\ and\ \bibinfo {author}
  {\bibfnamefont {Aaron}\ \bibnamefont {Courville}},\ }\href
  {http://www.deeplearningbook.org} {\enquote {\bibinfo {title} {Deep
  learning},}\ } (\bibinfo {year} {2016}),\ \bibinfo {note} {MIT
  Press}\BibitemShut {NoStop}%
\bibitem [{\citenamefont {Goodfellow}(2016)}]{goodfellow2016nips}%
  \BibitemOpen
  \bibfield  {author} {\bibinfo {author} {\bibfnamefont {Ian}\ \bibnamefont
  {Goodfellow}},\ }\bibfield  {title} {\enquote {\bibinfo {title} {Nips 2016
  tutorial: Generative adversarial networks},}\ }\href@noop {} {\bibfield
  {journal} {\bibinfo  {journal} {arXiv:1701.00160}\ } (\bibinfo {year}
  {2016})}\BibitemShut {NoStop}%
\bibitem [{\citenamefont {Ng}\ and\ \citenamefont
  {Jordan}(2002)}]{ng2002discriminative}%
  \BibitemOpen
  \bibfield  {author} {\bibinfo {author} {\bibfnamefont {Andrew~Y}\
  \bibnamefont {Ng}}\ and\ \bibinfo {author} {\bibfnamefont {Michael~I}\
  \bibnamefont {Jordan}},\ }\bibfield  {title} {\enquote {\bibinfo {title} {On
  discriminative vs. generative classifiers: A comparison of logistic
  regression and naive bayes},}\ }in\ \href@noop {} {\emph {\bibinfo
  {booktitle} {Advances in neural information processing systems}}}\ (\bibinfo
  {year} {2002})\ pp.\ \bibinfo {pages} {841--848}\BibitemShut {NoStop}%
\bibitem [{\citenamefont {Salakhutdinov}(2015)}]{Salakhutdinov-Review-2015}%
  \BibitemOpen
  \bibfield  {author} {\bibinfo {author} {\bibfnamefont {Ruslan}\ \bibnamefont
  {Salakhutdinov}},\ }\bibfield  {title} {\enquote {\bibinfo {title} {Learning
  deep generative models},}\ }\href {\doibase
  10.1146/annurev-statistics-010814-020120} {\bibfield  {journal} {\bibinfo
  {journal} {Annual Review of Statistics and Its Application}\ }\textbf
  {\bibinfo {volume} {2}},\ \bibinfo {pages} {361--385} (\bibinfo {year}
  {2015})}\BibitemShut {NoStop}%
\bibitem [{\citenamefont {Sinclair}\ and\ \citenamefont
  {Jerrum}(1989)}]{Sinclair-InfComp-1989}%
  \BibitemOpen
  \bibfield  {author} {\bibinfo {author} {\bibfnamefont {Alistair}\
  \bibnamefont {Sinclair}}\ and\ \bibinfo {author} {\bibfnamefont {Mark}\
  \bibnamefont {Jerrum}},\ }\bibfield  {title} {\enquote {\bibinfo {title}
  {Approximate counting, uniform generation and rapidly mixing markov
  chains},}\ }\href {\doibase 10.1016/0890-5401(89)90067-9} {\bibfield
  {journal} {\bibinfo  {journal} {Inf. Comput.}\ }\textbf {\bibinfo {volume}
  {82}},\ \bibinfo {pages} {93--133} (\bibinfo {year} {1989})}\BibitemShut
  {NoStop}%
\bibitem [{\citenamefont {Frigessi}\ \emph {et~al.}(1997)\citenamefont
  {Frigessi}, \citenamefont {Martinelli},\ and\ \citenamefont
  {Stander}}]{Frigessi-Biometrika-1997}%
  \BibitemOpen
  \bibfield  {author} {\bibinfo {author} {\bibfnamefont {Arnoldo}\ \bibnamefont
  {Frigessi}}, \bibinfo {author} {\bibfnamefont {Fabio}\ \bibnamefont
  {Martinelli}}, \ and\ \bibinfo {author} {\bibfnamefont {Julian}\ \bibnamefont
  {Stander}},\ }\bibfield  {title} {\enquote {\bibinfo {title} {{Computational
  complexity of Markov chain Monte Carlo methods for finite Markov random
  fields}},}\ }\href {\doibase 10.1093/biomet/84.1.1} {\bibfield  {journal}
  {\bibinfo  {journal} {Biometrika}\ }\textbf {\bibinfo {volume} {84}},\
  \bibinfo {pages} {1--18} (\bibinfo {year} {1997})}\BibitemShut {NoStop}%
\bibitem [{\citenamefont {Neven}\ \emph {et~al.}(2009)\citenamefont {Neven},
  \citenamefont {Denchev}, \citenamefont {Drew-Brook}, \citenamefont {Zhang},
  \citenamefont {Macready},\ and\ \citenamefont {Rose}}]{neven2009nips}%
  \BibitemOpen
  \bibfield  {author} {\bibinfo {author} {\bibfnamefont {Harmut}\ \bibnamefont
  {Neven}}, \bibinfo {author} {\bibfnamefont {Vasil~S}\ \bibnamefont
  {Denchev}}, \bibinfo {author} {\bibfnamefont {Marshall}\ \bibnamefont
  {Drew-Brook}}, \bibinfo {author} {\bibfnamefont {Jiayong}\ \bibnamefont
  {Zhang}}, \bibinfo {author} {\bibfnamefont {William~G}\ \bibnamefont
  {Macready}}, \ and\ \bibinfo {author} {\bibfnamefont {Geordie}\ \bibnamefont
  {Rose}},\ }\bibfield  {title} {\enquote {\bibinfo {title} {{Binary
  classification using hardware implementation of quantum annealing}},}\ }in\
  \href@noop {} {\emph {\bibinfo {booktitle} {{Demonstrations at NIPS-09, 24th
  Annual Conference on Neural Information Processing Systems}}}}\ (\bibinfo
  {year} {2009})\ pp.\ \bibinfo {pages} {1--17}\BibitemShut {NoStop}%
\bibitem [{\citenamefont {Bian}\ \emph {et~al.}(2010)\citenamefont {Bian},
  \citenamefont {Chudak}, \citenamefont {Macready},\ and\ \citenamefont
  {Rose}}]{bian2010ising}%
  \BibitemOpen
  \bibfield  {author} {\bibinfo {author} {\bibfnamefont {Zhengbing}\
  \bibnamefont {Bian}}, \bibinfo {author} {\bibfnamefont {Fabian}\ \bibnamefont
  {Chudak}}, \bibinfo {author} {\bibfnamefont {William~G}\ \bibnamefont
  {Macready}}, \ and\ \bibinfo {author} {\bibfnamefont {Geordie}\ \bibnamefont
  {Rose}},\ }\href@noop {} {\emph {\bibinfo {title} {{The Ising model: teaching
  an old problem new tricks}}}},\ \bibinfo {type} {Tech. Rep.}\ (\bibinfo
  {institution} {D-Wave Systems},\ \bibinfo {year} {2010})\BibitemShut
  {NoStop}%
\bibitem [{\citenamefont {Denil}\ and\ \citenamefont
  {De~Freitas}(2011)}]{Denil-2011}%
  \BibitemOpen
  \bibfield  {author} {\bibinfo {author} {\bibfnamefont {Misha}\ \bibnamefont
  {Denil}}\ and\ \bibinfo {author} {\bibfnamefont {Nando}\ \bibnamefont
  {De~Freitas}},\ }\bibfield  {title} {\enquote {\bibinfo {title} {Toward the
  implementation of a quantum {RBM}},}\ }\href@noop {} {\bibfield  {journal}
  {\bibinfo  {journal} {NIPS Deep Learning and Unsupervised Feature Learning
  Workshop}\ } (\bibinfo {year} {2011})}\BibitemShut {NoStop}%
\bibitem [{\citenamefont {Wiebe}\ \emph {et~al.}(2012)\citenamefont {Wiebe},
  \citenamefont {Braun},\ and\ \citenamefont {Lloyd}}]{wiebe2012quantum}%
  \BibitemOpen
  \bibfield  {author} {\bibinfo {author} {\bibfnamefont {Nathan}\ \bibnamefont
  {Wiebe}}, \bibinfo {author} {\bibfnamefont {Daniel}\ \bibnamefont {Braun}}, \
  and\ \bibinfo {author} {\bibfnamefont {Seth}\ \bibnamefont {Lloyd}},\
  }\bibfield  {title} {\enquote {\bibinfo {title} {Quantum algorithm for data
  fitting},}\ }\href@noop {} {\bibfield  {journal} {\bibinfo  {journal}
  {Physical review letters}\ }\textbf {\bibinfo {volume} {109}},\ \bibinfo
  {pages} {050505} (\bibinfo {year} {2012})}\BibitemShut {NoStop}%
\bibitem [{\citenamefont {Pudenz}\ and\ \citenamefont
  {Lidar}(2013)}]{Pudenz-QIP-2013}%
  \BibitemOpen
  \bibfield  {author} {\bibinfo {author} {\bibfnamefont {Kristen~L.}\
  \bibnamefont {Pudenz}}\ and\ \bibinfo {author} {\bibfnamefont {Daniel~A.}\
  \bibnamefont {Lidar}},\ }\bibfield  {title} {\enquote {\bibinfo {title}
  {Quantum adiabatic machine learning},}\ }\href {\doibase
  10.1007/s11128-012-0506-4} {\bibfield  {journal} {\bibinfo  {journal}
  {Quantum Information Processing}\ }\textbf {\bibinfo {volume} {12}},\
  \bibinfo {pages} {2027--2070} (\bibinfo {year} {2013})}\BibitemShut {NoStop}%
\bibitem [{\citenamefont {Lloyd}\ \emph {et~al.}(2013)\citenamefont {Lloyd},
  \citenamefont {Mohseni},\ and\ \citenamefont
  {Rebentrost}}]{Lloyd-arXiv-2013}%
  \BibitemOpen
  \bibfield  {author} {\bibinfo {author} {\bibfnamefont {Seth}\ \bibnamefont
  {Lloyd}}, \bibinfo {author} {\bibfnamefont {Masoud}\ \bibnamefont {Mohseni}},
  \ and\ \bibinfo {author} {\bibfnamefont {Patrick}\ \bibnamefont
  {Rebentrost}},\ }\bibfield  {title} {\enquote {\bibinfo {title} {Quantum
  algorithms for supervised and unsupervised machine learning},}\ }\href@noop
  {} {\bibfield  {journal} {\bibinfo  {journal} {arXiv:1307.0411}\ } (\bibinfo
  {year} {2013})}\BibitemShut {NoStop}%
\bibitem [{\citenamefont {Rebentrost}\ \emph {et~al.}(2014)\citenamefont
  {Rebentrost}, \citenamefont {Mohseni},\ and\ \citenamefont
  {Lloyd}}]{Rebentrost-PRL-2014}%
  \BibitemOpen
  \bibfield  {author} {\bibinfo {author} {\bibfnamefont {Patrick}\ \bibnamefont
  {Rebentrost}}, \bibinfo {author} {\bibfnamefont {Masoud}\ \bibnamefont
  {Mohseni}}, \ and\ \bibinfo {author} {\bibfnamefont {Seth}\ \bibnamefont
  {Lloyd}},\ }\bibfield  {title} {\enquote {\bibinfo {title} {Quantum support
  vector machine for big data classification},}\ }\href {\doibase
  10.1103/PhysRevLett.113.130503} {\bibfield  {journal} {\bibinfo  {journal}
  {Phys. Rev. Lett.}\ }\textbf {\bibinfo {volume} {113}},\ \bibinfo {pages}
  {130503} (\bibinfo {year} {2014})}\BibitemShut {NoStop}%
\bibitem [{\citenamefont {Wang}(2017)}]{wang2017quantum}%
  \BibitemOpen
  \bibfield  {author} {\bibinfo {author} {\bibfnamefont {Guoming}\ \bibnamefont
  {Wang}},\ }\bibfield  {title} {\enquote {\bibinfo {title} {Quantum algorithm
  for linear regression},}\ }\href@noop {} {\bibfield  {journal} {\bibinfo
  {journal} {Physical Review A}\ }\textbf {\bibinfo {volume} {96}},\ \bibinfo
  {pages} {012335} (\bibinfo {year} {2017})}\BibitemShut {NoStop}%
\bibitem [{\citenamefont {{Zhao}}\ \emph {et~al.}(2015)\citenamefont {{Zhao}},
  \citenamefont {{Fitzsimons}},\ and\ \citenamefont
  {{Fitzsimons}}}]{2015arXiv151203929Z}%
  \BibitemOpen
  \bibfield  {author} {\bibinfo {author} {\bibfnamefont {Z.}~\bibnamefont
  {{Zhao}}}, \bibinfo {author} {\bibfnamefont {J.~K.}\ \bibnamefont
  {{Fitzsimons}}}, \ and\ \bibinfo {author} {\bibfnamefont {J.~F.}\
  \bibnamefont {{Fitzsimons}}},\ }\bibfield  {title} {\enquote {\bibinfo
  {title} {{Quantum assisted Gaussian process regression}},}\ }\href@noop {}
  {\bibfield  {journal} {\bibinfo  {journal} {ArXiv e-prints}\ } (\bibinfo
  {year} {2015})},\ \Eprint {http://arxiv.org/abs/1512.03929} {arXiv:1512.03929
  [quant-ph]} \BibitemShut {NoStop}%
\bibitem [{\citenamefont {Lloyd}\ \emph {et~al.}(2014)\citenamefont {Lloyd},
  \citenamefont {Mohseni},\ and\ \citenamefont
  {Rebentrost}}]{Lloyd-NatPhys-2014}%
  \BibitemOpen
  \bibfield  {author} {\bibinfo {author} {\bibfnamefont {Seth}\ \bibnamefont
  {Lloyd}}, \bibinfo {author} {\bibfnamefont {Masoud}\ \bibnamefont {Mohseni}},
  \ and\ \bibinfo {author} {\bibfnamefont {Patrick}\ \bibnamefont
  {Rebentrost}},\ }\bibfield  {title} {\enquote {\bibinfo {title} {Quantum
  principal component analysis},}\ }\href@noop {} {\bibfield  {journal}
  {\bibinfo  {journal} {Nature Physics}\ }\textbf {\bibinfo {volume} {10}},\
  \bibinfo {pages} {631--633} (\bibinfo {year} {2014})}\BibitemShut {NoStop}%
\bibitem [{\citenamefont {Schuld}\ \emph {et~al.}(2016)\citenamefont {Schuld},
  \citenamefont {Sinayskiy},\ and\ \citenamefont
  {Petruccione}}]{schuld2016prediction}%
  \BibitemOpen
  \bibfield  {author} {\bibinfo {author} {\bibfnamefont {Maria}\ \bibnamefont
  {Schuld}}, \bibinfo {author} {\bibfnamefont {Ilya}\ \bibnamefont
  {Sinayskiy}}, \ and\ \bibinfo {author} {\bibfnamefont {Francesco}\
  \bibnamefont {Petruccione}},\ }\bibfield  {title} {\enquote {\bibinfo {title}
  {Prediction by linear regression on a quantum computer},}\ }\href@noop {}
  {\bibfield  {journal} {\bibinfo  {journal} {Physical Review A}\ }\textbf
  {\bibinfo {volume} {94}},\ \bibinfo {pages} {022342} (\bibinfo {year}
  {2016})}\BibitemShut {NoStop}%
\bibitem [{\citenamefont {Nathan~Wiebe}(2015)}]{Wiebe-arXiv-2015}%
  \BibitemOpen
  \bibfield  {author} {\bibinfo {author} {\bibfnamefont {Krysta M.~Svore}\
  \bibnamefont {Nathan~Wiebe}, \bibfnamefont {Ashish~Kapoor}},\ }\bibfield
  {title} {\enquote {\bibinfo {title} {Quantum deep learning},}\ }\href@noop {}
  {\bibfield  {journal} {\bibinfo  {journal} {arXiv:1412.3489}\ } (\bibinfo
  {year} {2015})}\BibitemShut {NoStop}%
\bibitem [{\citenamefont {Benedetti}\ \emph {et~al.}(2016)\citenamefont
  {Benedetti}, \citenamefont {Realpe-G\'omez}, \citenamefont {Biswas},\ and\
  \citenamefont {Perdomo-Ortiz}}]{Benedetti-2016}%
  \BibitemOpen
  \bibfield  {author} {\bibinfo {author} {\bibfnamefont {Marcello}\
  \bibnamefont {Benedetti}}, \bibinfo {author} {\bibfnamefont {John}\
  \bibnamefont {Realpe-G\'omez}}, \bibinfo {author} {\bibfnamefont {Rupak}\
  \bibnamefont {Biswas}}, \ and\ \bibinfo {author} {\bibfnamefont {Alejandro}\
  \bibnamefont {Perdomo-Ortiz}},\ }\bibfield  {title} {\enquote {\bibinfo
  {title} {Estimation of effective temperatures in quantum annealers for
  sampling applications: A case study with possible applications in deep
  learning},}\ }\href {\doibase 10.1103/PhysRevA.94.022308} {\bibfield
  {journal} {\bibinfo  {journal} {Phys. Rev. A}\ }\textbf {\bibinfo {volume}
  {94}},\ \bibinfo {pages} {022308} (\bibinfo {year} {2016})}\BibitemShut
  {NoStop}%
\bibitem [{\citenamefont {Aaronson}(2015)}]{Aaronson-2015}%
  \BibitemOpen
  \bibfield  {author} {\bibinfo {author} {\bibfnamefont {Scott}\ \bibnamefont
  {Aaronson}},\ }\bibfield  {title} {\enquote {\bibinfo {title} {Read the fine
  print},}\ }\href@noop {} {\bibfield  {journal} {\bibinfo  {journal} {Nature
  Physics}\ }\textbf {\bibinfo {volume} {11}},\ \bibinfo {pages} {291--293}
  (\bibinfo {year} {2015})},\ \bibinfo {note} {commentary}\BibitemShut
  {NoStop}%
\bibitem [{\citenamefont {Adachi}\ and\ \citenamefont
  {Henderson}(2015)}]{Adachi-arXiv-2015}%
  \BibitemOpen
  \bibfield  {author} {\bibinfo {author} {\bibfnamefont {Steven~H.}\
  \bibnamefont {Adachi}}\ and\ \bibinfo {author} {\bibfnamefont {Maxwell~P.}\
  \bibnamefont {Henderson}},\ }\bibfield  {title} {\enquote {\bibinfo {title}
  {Application of quantum annealing to training of deep neural networks},}\
  }\href@noop {} {\bibfield  {journal} {\bibinfo  {journal} {arXiv:1510.06356}\
  } (\bibinfo {year} {2015})}\BibitemShut {NoStop}%
\bibitem [{\citenamefont {Chancellor}\ \emph {et~al.}(2016)\citenamefont
  {Chancellor}, \citenamefont {Szoke}, \citenamefont {Vinci}, \citenamefont
  {Aeppli},\ and\ \citenamefont {Warburton}}]{chancellor2016maximum}%
  \BibitemOpen
  \bibfield  {author} {\bibinfo {author} {\bibfnamefont {Nicholas}\
  \bibnamefont {Chancellor}}, \bibinfo {author} {\bibfnamefont {Szilard}\
  \bibnamefont {Szoke}}, \bibinfo {author} {\bibfnamefont {Walter}\
  \bibnamefont {Vinci}}, \bibinfo {author} {\bibfnamefont {Gabriel}\
  \bibnamefont {Aeppli}}, \ and\ \bibinfo {author} {\bibfnamefont {Paul~A}\
  \bibnamefont {Warburton}},\ }\bibfield  {title} {\enquote {\bibinfo {title}
  {Maximum-entropy inference with a programmable annealer},}\ }\href@noop {}
  {\bibfield  {journal} {\bibinfo  {journal} {Scientific reports}\ }\textbf
  {\bibinfo {volume} {6}} (\bibinfo {year} {2016})}\BibitemShut {NoStop}%
\bibitem [{\citenamefont {{Mohammad H. Amin and Evgeny Andriyash and Jason
  Rolfe and Bohdan Kulchytskyy and Roger Melko}}(2016)}]{Amin-arXiv-2016}%
  \BibitemOpen
  \bibfield  {author} {\bibinfo {author} {\bibnamefont {{Mohammad H. Amin and
  Evgeny Andriyash and Jason Rolfe and Bohdan Kulchytskyy and Roger Melko}}},\
  }\bibfield  {title} {\enquote {\bibinfo {title} {{Quantum Boltzmann
  Machine}},}\ }\href@noop {} {\bibfield  {journal} {\bibinfo  {journal}
  {arXiv:1601.02036}\ } (\bibinfo {year} {2016})}\BibitemShut {NoStop}%
\bibitem [{\citenamefont {Kieferova}\ and\ \citenamefont
  {Wiebe}(2016)}]{kieferova2016tomography}%
  \BibitemOpen
  \bibfield  {author} {\bibinfo {author} {\bibfnamefont {Maria}\ \bibnamefont
  {Kieferova}}\ and\ \bibinfo {author} {\bibfnamefont {Nathan}\ \bibnamefont
  {Wiebe}},\ }\bibfield  {title} {\enquote {\bibinfo {title} {Tomography and
  generative data modeling via quantum boltzmann training},}\ }\href@noop {}
  {\bibfield  {journal} {\bibinfo  {journal} {arXiv preprint arXiv:1612.05204}\
  } (\bibinfo {year} {2016})}\BibitemShut {NoStop}%
\bibitem [{\citenamefont {Kerenidis}\ and\ \citenamefont
  {Prakash}(2016)}]{kerenidis2016quantum}%
  \BibitemOpen
  \bibfield  {author} {\bibinfo {author} {\bibfnamefont {Iordanis}\
  \bibnamefont {Kerenidis}}\ and\ \bibinfo {author} {\bibfnamefont {Anupam}\
  \bibnamefont {Prakash}},\ }\bibfield  {title} {\enquote {\bibinfo {title}
  {Quantum recommendation systems},}\ }\href@noop {} {\bibfield  {journal}
  {\bibinfo  {journal} {arXiv preprint arXiv:1603.08675}\ } (\bibinfo {year}
  {2016})}\BibitemShut {NoStop}%
\bibitem [{\citenamefont {Lamata}(2017)}]{lamata2017basic}%
  \BibitemOpen
  \bibfield  {author} {\bibinfo {author} {\bibfnamefont {Lucas}\ \bibnamefont
  {Lamata}},\ }\bibfield  {title} {\enquote {\bibinfo {title} {Basic protocols
  in quantum reinforcement learning with superconducting circuits},}\
  }\href@noop {} {\bibfield  {journal} {\bibinfo  {journal} {Scientific
  Reports}\ }\textbf {\bibinfo {volume} {7}} (\bibinfo {year}
  {2017})}\BibitemShut {NoStop}%
\bibitem [{\citenamefont {Alvarez-Rodriguez}\ \emph {et~al.}(2016)\citenamefont
  {Alvarez-Rodriguez}, \citenamefont {Lamata}, \citenamefont
  {Escandell-Montero}, \citenamefont {Mart{\'\i}n-Guerrero},\ and\
  \citenamefont {Solano}}]{alvarez2016quantum}%
  \BibitemOpen
  \bibfield  {author} {\bibinfo {author} {\bibfnamefont {Unai}\ \bibnamefont
  {Alvarez-Rodriguez}}, \bibinfo {author} {\bibfnamefont {Lucas}\ \bibnamefont
  {Lamata}}, \bibinfo {author} {\bibfnamefont {Pablo}\ \bibnamefont
  {Escandell-Montero}}, \bibinfo {author} {\bibfnamefont {Jos{\'e}~D}\
  \bibnamefont {Mart{\'\i}n-Guerrero}}, \ and\ \bibinfo {author} {\bibfnamefont
  {Enrique}\ \bibnamefont {Solano}},\ }\bibfield  {title} {\enquote {\bibinfo
  {title} {Quantum machine learning without measurements},}\ }\href@noop {}
  {\bibfield  {journal} {\bibinfo  {journal} {arXiv preprint arXiv:1612.05535}\
  } (\bibinfo {year} {2016})}\BibitemShut {NoStop}%
\bibitem [{\citenamefont {Wittek}\ and\ \citenamefont
  {Gogolin}(2017)}]{wittek2017quantum}%
  \BibitemOpen
  \bibfield  {author} {\bibinfo {author} {\bibfnamefont {Peter}\ \bibnamefont
  {Wittek}}\ and\ \bibinfo {author} {\bibfnamefont {Christian}\ \bibnamefont
  {Gogolin}},\ }\bibfield  {title} {\enquote {\bibinfo {title} {Quantum
  enhanced inference in markov logic networks},}\ }\href@noop {} {\bibfield
  {journal} {\bibinfo  {journal} {Scientific Reports}\ }\textbf {\bibinfo
  {volume} {7}} (\bibinfo {year} {2017})}\BibitemShut {NoStop}%
\bibitem [{\citenamefont {Potok}\ \emph {et~al.}(2017)\citenamefont {Potok},
  \citenamefont {Schuman}, \citenamefont {Young}, \citenamefont {Patton},
  \citenamefont {Spedalieri}, \citenamefont {Liu}, \citenamefont {Yao},
  \citenamefont {Rose},\ and\ \citenamefont {Chakma}}]{Potok2017}%
  \BibitemOpen
  \bibfield  {author} {\bibinfo {author} {\bibfnamefont {Thomas~E.}\
  \bibnamefont {Potok}}, \bibinfo {author} {\bibfnamefont {Catherine}\
  \bibnamefont {Schuman}}, \bibinfo {author} {\bibfnamefont {Steven~R.}\
  \bibnamefont {Young}}, \bibinfo {author} {\bibfnamefont {Robert~M.}\
  \bibnamefont {Patton}}, \bibinfo {author} {\bibfnamefont {Federico}\
  \bibnamefont {Spedalieri}}, \bibinfo {author} {\bibfnamefont {Jeremy}\
  \bibnamefont {Liu}}, \bibinfo {author} {\bibfnamefont {Ke-Thia}\ \bibnamefont
  {Yao}}, \bibinfo {author} {\bibfnamefont {Garrett}\ \bibnamefont {Rose}}, \
  and\ \bibinfo {author} {\bibfnamefont {Gangotree}\ \bibnamefont {Chakma}},\
  }\bibfield  {title} {\enquote {\bibinfo {title} {A study of complex deep
  learning networks on high performance, neuromorphic, and quantum
  computers},}\ }\href@noop {} {\bibfield  {journal} {\bibinfo  {journal}
  {arXiv:1703.05364}\ } (\bibinfo {year} {2017})}\BibitemShut {NoStop}%
\bibitem [{\citenamefont {Schuld}\ \emph {et~al.}(2015)\citenamefont {Schuld},
  \citenamefont {Sinayskiy},\ and\ \citenamefont
  {Petruccione}}]{Schuld-QML-2015}%
  \BibitemOpen
  \bibfield  {author} {\bibinfo {author} {\bibfnamefont {Maria}\ \bibnamefont
  {Schuld}}, \bibinfo {author} {\bibfnamefont {Ilya}\ \bibnamefont
  {Sinayskiy}}, \ and\ \bibinfo {author} {\bibfnamefont {Francesco}\
  \bibnamefont {Petruccione}},\ }\bibfield  {title} {\enquote {\bibinfo {title}
  {An introduction to quantum machine learning},}\ }\href@noop {} {\bibfield
  {journal} {\bibinfo  {journal} {Contemporary Physics}\ }\textbf {\bibinfo
  {volume} {56}},\ \bibinfo {pages} {172--185} (\bibinfo {year}
  {2015})}\BibitemShut {NoStop}%
\bibitem [{\citenamefont {Romero}\ \emph {et~al.}(2017)\citenamefont {Romero},
  \citenamefont {Olson},\ and\ \citenamefont {Aspuru-Guzik}}]{Romero2017}%
  \BibitemOpen
  \bibfield  {author} {\bibinfo {author} {\bibfnamefont {Jonathan}\
  \bibnamefont {Romero}}, \bibinfo {author} {\bibfnamefont {Jonathan~P}\
  \bibnamefont {Olson}}, \ and\ \bibinfo {author} {\bibfnamefont {Alan}\
  \bibnamefont {Aspuru-Guzik}},\ }\bibfield  {title} {\enquote {\bibinfo
  {title} {Quantum autoencoders for efficient compression of quantum data},}\
  }\href@noop {} {\bibfield  {journal} {\bibinfo  {journal} {Quantum Sci.
  Technol.}\ }\textbf {\bibinfo {volume} {2}},\ \bibinfo {pages} {045001}
  (\bibinfo {year} {2017})}\BibitemShut {NoStop}%
\bibitem [{\citenamefont {Adcock}\ \emph {et~al.}(2015)\citenamefont {Adcock},
  \citenamefont {Allen}, \citenamefont {Day}, \citenamefont {Frick},
  \citenamefont {Hinchliff}, \citenamefont {Johnson}, \citenamefont
  {Morley-Short}, \citenamefont {Pallister}, \citenamefont {Price},\ and\
  \citenamefont {Stanisic}}]{adcock2015advances}%
  \BibitemOpen
  \bibfield  {author} {\bibinfo {author} {\bibfnamefont {Jeremy}\ \bibnamefont
  {Adcock}}, \bibinfo {author} {\bibfnamefont {Euan}\ \bibnamefont {Allen}},
  \bibinfo {author} {\bibfnamefont {Matthew}\ \bibnamefont {Day}}, \bibinfo
  {author} {\bibfnamefont {Stefan}\ \bibnamefont {Frick}}, \bibinfo {author}
  {\bibfnamefont {Janna}\ \bibnamefont {Hinchliff}}, \bibinfo {author}
  {\bibfnamefont {Mack}\ \bibnamefont {Johnson}}, \bibinfo {author}
  {\bibfnamefont {Sam}\ \bibnamefont {Morley-Short}}, \bibinfo {author}
  {\bibfnamefont {Sam}\ \bibnamefont {Pallister}}, \bibinfo {author}
  {\bibfnamefont {Alasdair}\ \bibnamefont {Price}}, \ and\ \bibinfo {author}
  {\bibfnamefont {Stasja}\ \bibnamefont {Stanisic}},\ }\bibfield  {title}
  {\enquote {\bibinfo {title} {Advances in quantum machine learning},}\
  }\href@noop {} {\bibfield  {journal} {\bibinfo  {journal} {arXiv preprint
  arXiv:1512.02900}\ } (\bibinfo {year} {2015})}\BibitemShut {NoStop}%
\bibitem [{\citenamefont {Biamonte}\ \emph {et~al.}(2016)\citenamefont
  {Biamonte}, \citenamefont {Wittek}, \citenamefont {Pancotti}, \citenamefont
  {Rebentrost}, \citenamefont {Wiebe},\ and\ \citenamefont
  {Lloyd}}]{biamonte2016quantum}%
  \BibitemOpen
  \bibfield  {author} {\bibinfo {author} {\bibfnamefont {Jacob}\ \bibnamefont
  {Biamonte}}, \bibinfo {author} {\bibfnamefont {Peter}\ \bibnamefont
  {Wittek}}, \bibinfo {author} {\bibfnamefont {Nicola}\ \bibnamefont
  {Pancotti}}, \bibinfo {author} {\bibfnamefont {Patrick}\ \bibnamefont
  {Rebentrost}}, \bibinfo {author} {\bibfnamefont {Nathan}\ \bibnamefont
  {Wiebe}}, \ and\ \bibinfo {author} {\bibfnamefont {Seth}\ \bibnamefont
  {Lloyd}},\ }\bibfield  {title} {\enquote {\bibinfo {title} {Quantum machine
  learning},}\ }\href@noop {} {\bibfield  {journal} {\bibinfo  {journal} {arXiv
  preprint arXiv:1611.09347}\ } (\bibinfo {year} {2016})}\BibitemShut {NoStop}%
\bibitem [{\citenamefont {{Ciliberto}}\ \emph {et~al.}(2017)\citenamefont
  {{Ciliberto}}, \citenamefont {{Herbster}}, \citenamefont {{Davide Ialongo}},
  \citenamefont {{Pontil}}, \citenamefont {{Rocchetto}}, \citenamefont
  {{Severini}},\ and\ \citenamefont {{Wossnig}}}]{2017arXiv170708561C}%
  \BibitemOpen
  \bibfield  {author} {\bibinfo {author} {\bibfnamefont {C.}~\bibnamefont
  {{Ciliberto}}}, \bibinfo {author} {\bibfnamefont {M.}~\bibnamefont
  {{Herbster}}}, \bibinfo {author} {\bibfnamefont {A.}~\bibnamefont {{Davide
  Ialongo}}}, \bibinfo {author} {\bibfnamefont {M.}~\bibnamefont {{Pontil}}},
  \bibinfo {author} {\bibfnamefont {A.}~\bibnamefont {{Rocchetto}}}, \bibinfo
  {author} {\bibfnamefont {S.}~\bibnamefont {{Severini}}}, \ and\ \bibinfo
  {author} {\bibfnamefont {L.}~\bibnamefont {{Wossnig}}},\ }\bibfield  {title}
  {\enquote {\bibinfo {title} {{Quantum machine learning: a classical
  perspective}},}\ }\href@noop {} {\bibfield  {journal} {\bibinfo  {journal}
  {ArXiv e-prints}\ } (\bibinfo {year} {2017})},\ \Eprint
  {http://arxiv.org/abs/1707.08561} {arXiv:1707.08561 [quant-ph]} \BibitemShut
  {NoStop}%
\bibitem [{\citenamefont {Perdomo-Ortiz}\ \emph
  {et~al.}(2017{\natexlab{a}})\citenamefont {Perdomo-Ortiz}, \citenamefont
  {Benedetti}, \citenamefont {Realpe-G\'omez},\ and\ \citenamefont
  {Biswas}}]{PerdomoOrtiz2017}%
  \BibitemOpen
  \bibfield  {author} {\bibinfo {author} {\bibfnamefont {Alejandro}\
  \bibnamefont {Perdomo-Ortiz}}, \bibinfo {author} {\bibfnamefont {Marcello}\
  \bibnamefont {Benedetti}}, \bibinfo {author} {\bibfnamefont {John}\
  \bibnamefont {Realpe-G\'omez}}, \ and\ \bibinfo {author} {\bibfnamefont
  {Rupak}\ \bibnamefont {Biswas}},\ }\bibfield  {title} {\enquote {\bibinfo
  {title} {Opportunities and challenges for quantum-assisted machine learning
  in near-term quantum computers},}\ }\href@noop {} {\bibfield  {journal}
  {\bibinfo  {journal} {arXiv:1708.09757}\ } (\bibinfo {year}
  {2017}{\natexlab{a}})}\BibitemShut {NoStop}%
\bibitem [{\citenamefont {Benedetti}\ \emph {et~al.}(2017)\citenamefont
  {Benedetti}, \citenamefont {Realpe-G\'omez},\ and\ \citenamefont
  {Perdomo-Ortiz}}]{Benedetti2017b}%
  \BibitemOpen
  \bibfield  {author} {\bibinfo {author} {\bibfnamefont {Marcello}\
  \bibnamefont {Benedetti}}, \bibinfo {author} {\bibfnamefont {John}\
  \bibnamefont {Realpe-G\'omez}}, \ and\ \bibinfo {author} {\bibfnamefont
  {Alejandro}\ \bibnamefont {Perdomo-Ortiz}},\ }\bibfield  {title} {\enquote
  {\bibinfo {title} {Quantum-assisted helmholtz machines: A quantum-classical
  deep learning framework for industrial datasets in near-term devices},}\
  }\href@noop {} {\bibfield  {journal} {\bibinfo  {journal} {arXiv:1708.09784}\
  } (\bibinfo {year} {2017})}\BibitemShut {NoStop}%
\bibitem [{\citenamefont {Gu}\ \emph {et~al.}(2012)\citenamefont {Gu},
  \citenamefont {Wiesner}, \citenamefont {Rieper},\ and\ \citenamefont
  {Vedral}}]{Gu2012}%
  \BibitemOpen
  \bibfield  {author} {\bibinfo {author} {\bibfnamefont {Mile}\ \bibnamefont
  {Gu}}, \bibinfo {author} {\bibfnamefont {Karoline}\ \bibnamefont {Wiesner}},
  \bibinfo {author} {\bibfnamefont {Elisabeth}\ \bibnamefont {Rieper}}, \ and\
  \bibinfo {author} {\bibfnamefont {Vlatko}\ \bibnamefont {Vedral}},\
  }\bibfield  {title} {\enquote {\bibinfo {title} {Quantum mechanics can reduce
  the complexity of classical models},}\ }\href@noop {} {\bibfield  {journal}
  {\bibinfo  {journal} {Nature Communications}\ }\textbf {\bibinfo {volume}
  {3}},\ \bibinfo {pages} {762} (\bibinfo {year} {2012})}\BibitemShut {NoStop}%
\bibitem [{\citenamefont {Finnila}\ \emph {et~al.}(1994)\citenamefont
  {Finnila}, \citenamefont {Gomez}, \citenamefont {Sebenik}, \citenamefont
  {Stenson},\ and\ \citenamefont {Doll}}]{finnila1994quantum}%
  \BibitemOpen
  \bibfield  {author} {\bibinfo {author} {\bibfnamefont {AB}~\bibnamefont
  {Finnila}}, \bibinfo {author} {\bibfnamefont {MA}~\bibnamefont {Gomez}},
  \bibinfo {author} {\bibfnamefont {C}~\bibnamefont {Sebenik}}, \bibinfo
  {author} {\bibfnamefont {C}~\bibnamefont {Stenson}}, \ and\ \bibinfo {author}
  {\bibfnamefont {JD}~\bibnamefont {Doll}},\ }\bibfield  {title} {\enquote
  {\bibinfo {title} {{Quantum annealing: a new method for minimizing
  multidimensional functions}},}\ }\href@noop {} {\bibfield  {journal}
  {\bibinfo  {journal} {{Chemical Physics Letters}}\ }\textbf {\bibinfo
  {volume} {219}},\ \bibinfo {pages} {343--348} (\bibinfo {year}
  {1994})}\BibitemShut {NoStop}%
\bibitem [{\citenamefont {Kadowaki}\ and\ \citenamefont
  {Nishimori}(1998)}]{kadowaki_quantum_1998}%
  \BibitemOpen
  \bibfield  {author} {\bibinfo {author} {\bibfnamefont {Tadashi}\ \bibnamefont
  {Kadowaki}}\ and\ \bibinfo {author} {\bibfnamefont {Hidetoshi}\ \bibnamefont
  {Nishimori}},\ }\bibfield  {title} {\enquote {\bibinfo {title} {Quantum
  annealing in the transverse ising model},}\ }\href {\doibase
  10.1103/PhysRevE.58.5355} {\bibfield  {journal} {\bibinfo  {journal} {Phys.
  Rev. E.}\ }\textbf {\bibinfo {volume} {58}},\ \bibinfo {pages} {5355}
  (\bibinfo {year} {1998})}\BibitemShut {NoStop}%
\bibitem [{\citenamefont {Farhi}\ \emph {et~al.}(2001)\citenamefont {Farhi},
  \citenamefont {Goldstone}, \citenamefont {Gutmann}, \citenamefont {Lapan},
  \citenamefont {Lundgren},\ and\ \citenamefont {Preda}}]{Farhi2001}%
  \BibitemOpen
  \bibfield  {author} {\bibinfo {author} {\bibfnamefont {Edward}\ \bibnamefont
  {Farhi}}, \bibinfo {author} {\bibfnamefont {Jeffrey}\ \bibnamefont
  {Goldstone}}, \bibinfo {author} {\bibfnamefont {Sam}\ \bibnamefont
  {Gutmann}}, \bibinfo {author} {\bibfnamefont {Joshua}\ \bibnamefont {Lapan}},
  \bibinfo {author} {\bibfnamefont {Andrew}\ \bibnamefont {Lundgren}}, \ and\
  \bibinfo {author} {\bibfnamefont {Daniel}\ \bibnamefont {Preda}},\ }\bibfield
   {title} {\enquote {\bibinfo {title} {A quantum adiabatic evolution algorithm
  applied to random instances of an {NP-Complete} problem},}\ }\href {\doibase
  10.1126/science.1057726} {\bibfield  {journal} {\bibinfo  {journal}
  {Science}\ }\textbf {\bibinfo {volume} {292}},\ \bibinfo {pages} {472--475}
  (\bibinfo {year} {2001})}\BibitemShut {NoStop}%
\bibitem [{\citenamefont {Gaitan}\ and\ \citenamefont
  {Clark}(2012)}]{Gaitan2012}%
  \BibitemOpen
  \bibfield  {author} {\bibinfo {author} {\bibfnamefont {Frank}\ \bibnamefont
  {Gaitan}}\ and\ \bibinfo {author} {\bibfnamefont {Lane}\ \bibnamefont
  {Clark}},\ }\bibfield  {title} {\enquote {\bibinfo {title} {Ramsey numbers
  and adiabatic quantum computing},}\ }\href {\doibase
  10.1103/PhysRevLett.108.010501} {\bibfield  {journal} {\bibinfo  {journal}
  {Phys. Rev. Lett.}\ }\textbf {\bibinfo {volume} {108}},\ \bibinfo {pages}
  {010501} (\bibinfo {year} {2012})}\BibitemShut {NoStop}%
\bibitem [{\citenamefont {Perdomo-Ortiz}\ \emph {et~al.}(2012)\citenamefont
  {Perdomo-Ortiz}, \citenamefont {Dickson}, \citenamefont {Drew-Brook},
  \citenamefont {Rose},\ and\ \citenamefont
  {Aspuru-Guzik}}]{PerdomoOrtiz2012_LPF}%
  \BibitemOpen
  \bibfield  {author} {\bibinfo {author} {\bibfnamefont {A.}~\bibnamefont
  {Perdomo-Ortiz}}, \bibinfo {author} {\bibfnamefont {N.}~\bibnamefont
  {Dickson}}, \bibinfo {author} {\bibfnamefont {M.}~\bibnamefont {Drew-Brook}},
  \bibinfo {author} {\bibfnamefont {G.}~\bibnamefont {Rose}}, \ and\ \bibinfo
  {author} {\bibfnamefont {A.}~\bibnamefont {Aspuru-Guzik}},\ }\bibfield
  {title} {\enquote {\bibinfo {title} {Finding low-energy conformations of
  lattice protein models by quantum annealing},}\ }\href@noop {} {\bibfield
  {journal} {\bibinfo  {journal} {Sci. Rep.}\ }\textbf {\bibinfo {volume}
  {2}},\ \bibinfo {pages} {571} (\bibinfo {year} {2012})}\BibitemShut {NoStop}%
\bibitem [{\citenamefont {Bian}\ \emph {et~al.}(2014)\citenamefont {Bian},
  \citenamefont {Chudak}, \citenamefont {Israel}, \citenamefont {Lackey},
  \citenamefont {Macready},\ and\ \citenamefont {Roy}}]{Bian2014}%
  \BibitemOpen
  \bibfield  {author} {\bibinfo {author} {\bibfnamefont {Zhengbing}\
  \bibnamefont {Bian}}, \bibinfo {author} {\bibfnamefont {Fabian}\ \bibnamefont
  {Chudak}}, \bibinfo {author} {\bibfnamefont {Robert}\ \bibnamefont {Israel}},
  \bibinfo {author} {\bibfnamefont {Brad}\ \bibnamefont {Lackey}}, \bibinfo
  {author} {\bibfnamefont {William~G}\ \bibnamefont {Macready}}, \ and\
  \bibinfo {author} {\bibfnamefont {Aidan}\ \bibnamefont {Roy}},\ }\bibfield
  {title} {\enquote {\bibinfo {title} {Discrete optimization using quantum
  annealing on sparse ising models},}\ }\href {\doibase
  10.3389/fphy.2014.00056} {\bibfield  {journal} {\bibinfo  {journal}
  {Frontiers in Physics}\ }\textbf {\bibinfo {volume} {2}} (\bibinfo {year}
  {2014}),\ 10.3389/fphy.2014.00056}\BibitemShut {NoStop}%
\bibitem [{\citenamefont {O'Gorman}\ \emph {et~al.}(2015)\citenamefont
  {O'Gorman}, \citenamefont {Babbush}, \citenamefont {Perdomo-Ortiz},
  \citenamefont {Aspuru-Guzik},\ and\ \citenamefont
  {Smelyanskiy}}]{OGorman-EPJST-2015}%
  \BibitemOpen
  \bibfield  {author} {\bibinfo {author} {\bibfnamefont {B.}~\bibnamefont
  {O'Gorman}}, \bibinfo {author} {\bibfnamefont {R.}~\bibnamefont {Babbush}},
  \bibinfo {author} {\bibfnamefont {A.}~\bibnamefont {Perdomo-Ortiz}}, \bibinfo
  {author} {\bibfnamefont {A.}~\bibnamefont {Aspuru-Guzik}}, \ and\ \bibinfo
  {author} {\bibfnamefont {V.}~\bibnamefont {Smelyanskiy}},\ }\bibfield
  {title} {\enquote {\bibinfo {title} {Bayesian network structure learning
  using quantum annealing},}\ }\href {\doibase 10.1140/epjst/e2015-02349-9}
  {\bibfield  {journal} {\bibinfo  {journal} {The European Physical Journal
  Special Topics}\ }\textbf {\bibinfo {volume} {224}},\ \bibinfo {pages}
  {163--188} (\bibinfo {year} {2015})}\BibitemShut {NoStop}%
\bibitem [{\citenamefont {Rieffel}\ \emph {et~al.}(2015)\citenamefont
  {Rieffel}, \citenamefont {Venturelli}, \citenamefont {O'Gorman},
  \citenamefont {Do}, \citenamefont {Prystay},\ and\ \citenamefont
  {Smelyanskiy}}]{RieffelQIP2015}%
  \BibitemOpen
  \bibfield  {author} {\bibinfo {author} {\bibfnamefont {Eleanor~G.}\
  \bibnamefont {Rieffel}}, \bibinfo {author} {\bibfnamefont {Davide}\
  \bibnamefont {Venturelli}}, \bibinfo {author} {\bibfnamefont {Bryan}\
  \bibnamefont {O'Gorman}}, \bibinfo {author} {\bibfnamefont {Minh~B.}\
  \bibnamefont {Do}}, \bibinfo {author} {\bibfnamefont {Elicia~M.}\
  \bibnamefont {Prystay}}, \ and\ \bibinfo {author} {\bibfnamefont {Vadim~N.}\
  \bibnamefont {Smelyanskiy}},\ }\bibfield  {title} {\enquote {\bibinfo {title}
  {A case study in programming a quantum annealer for hard operational planning
  problems},}\ }\href {\doibase 10.1007/s11128-014-0892-x} {\bibfield
  {journal} {\bibinfo  {journal} {Quantum Information Processing}\ }\textbf
  {\bibinfo {volume} {14}},\ \bibinfo {pages} {1--36} (\bibinfo {year}
  {2015})}\BibitemShut {NoStop}%
\bibitem [{\citenamefont {Perdomo-Ortiz}\ \emph
  {et~al.}(2015{\natexlab{a}})\citenamefont {Perdomo-Ortiz}, \citenamefont
  {Fluegemann}, \citenamefont {Narasimhan}, \citenamefont {Biswas},\ and\
  \citenamefont {Smelyanskiy}}]{PerdomoOrtiz_EPJST2015}%
  \BibitemOpen
  \bibfield  {author} {\bibinfo {author} {\bibfnamefont {A.}~\bibnamefont
  {Perdomo-Ortiz}}, \bibinfo {author} {\bibfnamefont {J.}~\bibnamefont
  {Fluegemann}}, \bibinfo {author} {\bibfnamefont {S.}~\bibnamefont
  {Narasimhan}}, \bibinfo {author} {\bibfnamefont {R.}~\bibnamefont {Biswas}},
  \ and\ \bibinfo {author} {\bibfnamefont {V.~N.}\ \bibnamefont
  {Smelyanskiy}},\ }\bibfield  {title} {\enquote {\bibinfo {title} {A quantum
  annealing approach for fault detection and diagnosis of graph-based
  systems},}\ }\href@noop {} {\bibfield  {journal} {\bibinfo  {journal} {Eur.
  Phys. J. Special Topics}\ }\textbf {\bibinfo {volume} {224}},\ \bibinfo
  {pages} {131--148} (\bibinfo {year} {2015}{\natexlab{a}})}\BibitemShut
  {NoStop}%
\bibitem [{\citenamefont {Perdomo-Ortiz}\ \emph
  {et~al.}(2015{\natexlab{b}})\citenamefont {Perdomo-Ortiz}, \citenamefont
  {Fluegemann}, \citenamefont {Biswas},\ and\ \citenamefont
  {Smelyanskiy}}]{perdomo2015performance}%
  \BibitemOpen
  \bibfield  {author} {\bibinfo {author} {\bibfnamefont {Alejandro}\
  \bibnamefont {Perdomo-Ortiz}}, \bibinfo {author} {\bibfnamefont {Joseph}\
  \bibnamefont {Fluegemann}}, \bibinfo {author} {\bibfnamefont {Rupak}\
  \bibnamefont {Biswas}}, \ and\ \bibinfo {author} {\bibfnamefont {Vadim~N}\
  \bibnamefont {Smelyanskiy}},\ }\bibfield  {title} {\enquote {\bibinfo {title}
  {{A performance estimator for quantum annealers: gauge selection and
  parameter setting}},}\ }\href@noop {} {\bibfield  {journal} {\bibinfo
  {journal} {arXiv:1503.01083}\ } (\bibinfo {year}
  {2015}{\natexlab{b}})}\BibitemShut {NoStop}%
\bibitem [{\citenamefont {Venturelli}\ \emph
  {et~al.}(2015{\natexlab{a}})\citenamefont {Venturelli}, \citenamefont
  {Marchand},\ and\ \citenamefont {Rojo}}]{Venturelli-JobShop-arXiv-2015}%
  \BibitemOpen
  \bibfield  {author} {\bibinfo {author} {\bibfnamefont {Davide}\ \bibnamefont
  {Venturelli}}, \bibinfo {author} {\bibfnamefont {Dominic~J.J.}\ \bibnamefont
  {Marchand}}, \ and\ \bibinfo {author} {\bibfnamefont {Galo}\ \bibnamefont
  {Rojo}},\ }\bibfield  {title} {\enquote {\bibinfo {title} {Quantum annealing
  implementation of job-shop scheduling},}\ }\href@noop {} {\bibfield
  {journal} {\bibinfo  {journal} {arXiv:1506.08479}\ } (\bibinfo {year}
  {2015}{\natexlab{a}})}\BibitemShut {NoStop}%
\bibitem [{\citenamefont {Perdomo-Ortiz}\ \emph
  {et~al.}(2017{\natexlab{b}})\citenamefont {Perdomo-Ortiz}, \citenamefont
  {Feldman}, \citenamefont {Ozaeta}, \citenamefont {Isakov}, \citenamefont
  {Zhu}, \citenamefont {O'Gorman}, \citenamefont {Katzgraber}, \citenamefont
  {Diedrich}, \citenamefont {Neven}, \citenamefont {de~Kleer}, \citenamefont
  {Lackey},\ and\ \citenamefont {Biswas}}]{PerdomoOrtiz2017a}%
  \BibitemOpen
  \bibfield  {author} {\bibinfo {author} {\bibfnamefont {Alejandro}\
  \bibnamefont {Perdomo-Ortiz}}, \bibinfo {author} {\bibfnamefont {Alexander}\
  \bibnamefont {Feldman}}, \bibinfo {author} {\bibfnamefont {Asier}\
  \bibnamefont {Ozaeta}}, \bibinfo {author} {\bibfnamefont {Sergei~V.}\
  \bibnamefont {Isakov}}, \bibinfo {author} {\bibfnamefont {Zheng}\
  \bibnamefont {Zhu}}, \bibinfo {author} {\bibfnamefont {Bryan}\ \bibnamefont
  {O'Gorman}}, \bibinfo {author} {\bibfnamefont {Helmut~G.}\ \bibnamefont
  {Katzgraber}}, \bibinfo {author} {\bibfnamefont {Alexander}\ \bibnamefont
  {Diedrich}}, \bibinfo {author} {\bibfnamefont {Hartmut}\ \bibnamefont
  {Neven}}, \bibinfo {author} {\bibfnamefont {Johan}\ \bibnamefont {de~Kleer}},
  \bibinfo {author} {\bibfnamefont {Brad}\ \bibnamefont {Lackey}}, \ and\
  \bibinfo {author} {\bibfnamefont {Rupak}\ \bibnamefont {Biswas}},\ }\bibfield
   {title} {\enquote {\bibinfo {title} {On the readiness of quantum
  optimization machines for industrial applications},}\ }\href@noop {}
  {\bibfield  {journal} {\bibinfo  {journal} {arXiv:1708.09780}\ } (\bibinfo
  {year} {2017}{\natexlab{b}})}\BibitemShut {NoStop}%
\bibitem [{\citenamefont {Raymond}\ \emph {et~al.}(2016)\citenamefont
  {Raymond}, \citenamefont {Yarkoni},\ and\ \citenamefont
  {Andriyash}}]{Raymond-DWave-2016}%
  \BibitemOpen
  \bibfield  {author} {\bibinfo {author} {\bibfnamefont {Jack}\ \bibnamefont
  {Raymond}}, \bibinfo {author} {\bibfnamefont {Sheir}\ \bibnamefont
  {Yarkoni}}, \ and\ \bibinfo {author} {\bibfnamefont {Evgeny}\ \bibnamefont
  {Andriyash}},\ }\bibfield  {title} {\enquote {\bibinfo {title} {{Global
  warming: Temperature estimation in annealers}},}\ }\href@noop {} {\bibfield
  {journal} {\bibinfo  {journal} {arXiv:1606.00919}\ } (\bibinfo {year}
  {2016})}\BibitemShut {NoStop}%
\bibitem [{\citenamefont {Amin}(2015)}]{Amin-arXiv-2015}%
  \BibitemOpen
  \bibfield  {author} {\bibinfo {author} {\bibfnamefont {Mohammad~H.}\
  \bibnamefont {Amin}},\ }\bibfield  {title} {\enquote {\bibinfo {title}
  {Searching for quantum speedup in quasistatic quantum annealers},}\ }\href
  {\doibase 10.1103/PhysRevA.92.052323} {\bibfield  {journal} {\bibinfo
  {journal} {Phys. Rev. A}\ }\textbf {\bibinfo {volume} {92}},\ \bibinfo
  {pages} {052323} (\bibinfo {year} {2015})}\BibitemShut {NoStop}%
\bibitem [{\citenamefont {Benedetti}(2015)}]{Marc}%
  \BibitemOpen
  \bibfield  {author} {\bibinfo {author} {\bibfnamefont {Marcello}\
  \bibnamefont {Benedetti}},\ }\emph {\bibinfo {title} {{Exploring Quantum
  Annealing for Data Mining}}},\ \href@noop {} {Master's thesis},\ \bibinfo
  {school} {Universit{\'e} Lumi{\`e}re Lyon 2}, \bibinfo {address} {France}
  (\bibinfo {year} {2015})\BibitemShut {NoStop}%
\bibitem [{\citenamefont {Dumoulin}\ \emph {et~al.}(2014)\citenamefont
  {Dumoulin}, \citenamefont {Goodfellow}, \citenamefont {Courville},\ and\
  \citenamefont {Bengio}}]{Dumolin-2014}%
  \BibitemOpen
  \bibfield  {author} {\bibinfo {author} {\bibfnamefont {V.}~\bibnamefont
  {Dumoulin}}, \bibinfo {author} {\bibfnamefont {I.~J.}\ \bibnamefont
  {Goodfellow}}, \bibinfo {author} {\bibfnamefont {A.~C.}\ \bibnamefont
  {Courville}}, \ and\ \bibinfo {author} {\bibfnamefont {Y.}~\bibnamefont
  {Bengio}},\ }\bibfield  {title} {\enquote {\bibinfo {title} {On the
  challenges of physical implementations of {RBM}s},}\ }in\ \href@noop {}
  {\emph {\bibinfo {booktitle} {Proceedings of the Twenty-Eighth {AAAI}
  Conference on Artificial Intelligence, July 27 -31, 2014, Qu{\'{e}}bec City,
  Qu{\'{e}}bec, Canada.}}}\ (\bibinfo {year} {2014})\ pp.\ \bibinfo {pages}
  {1199--1205}\BibitemShut {NoStop}%
\bibitem [{\citenamefont {Schneidman}\ \emph {et~al.}(2006)\citenamefont
  {Schneidman}, \citenamefont {Berry}, \citenamefont {Segev},\ and\
  \citenamefont {Bialek}}]{Schneidman-Nature-2006}%
  \BibitemOpen
  \bibfield  {author} {\bibinfo {author} {\bibfnamefont {Elad}\ \bibnamefont
  {Schneidman}}, \bibinfo {author} {\bibfnamefont {Michael~J.}\ \bibnamefont
  {Berry}}, \bibinfo {author} {\bibfnamefont {Ronen}\ \bibnamefont {Segev}}, \
  and\ \bibinfo {author} {\bibfnamefont {William}\ \bibnamefont {Bialek}},\
  }\bibfield  {title} {\enquote {\bibinfo {title} {Weak pairwise correlations
  imply strongly correlated network states in a neural population},}\ }\href
  {\doibase 10.1038/nature04701} {\bibfield  {journal} {\bibinfo  {journal}
  {Nature}\ }\textbf {\bibinfo {volume} {440}},\ \bibinfo {pages} {1007--1012}
  (\bibinfo {year} {2006})}\BibitemShut {NoStop}%
\bibitem [{\citenamefont {Ricci-Tersenghi}(2012)}]{Ricci-Tersenghi-JSTAT-2012}%
  \BibitemOpen
  \bibfield  {author} {\bibinfo {author} {\bibfnamefont {Federico}\
  \bibnamefont {Ricci-Tersenghi}},\ }\bibfield  {title} {\enquote {\bibinfo
  {title} {The bethe approximation for solving the inverse ising problem: a
  comparison with other inference methods},}\ }\href@noop {} {\bibfield
  {journal} {\bibinfo  {journal} {Journal of Statistical Mechanics: Theory and
  Experiment}\ }\textbf {\bibinfo {volume} {2012}},\ \bibinfo {pages} {P08015}
  (\bibinfo {year} {2012})}\BibitemShut {NoStop}%
\bibitem [{\citenamefont {Lichman}(2013)}]{Lichman:2013}%
  \BibitemOpen
  \bibfield  {author} {\bibinfo {author} {\bibfnamefont {M.}~\bibnamefont
  {Lichman}},\ }\href {http://archive.ics.uci.edu/ml} {\enquote {\bibinfo
  {title} {{UCI} machine learning repository},}\ } (\bibinfo {year}
  {2013})\BibitemShut {NoStop}%
\bibitem [{\citenamefont {MacKay}(2002)}]{MacKay-book-2002}%
  \BibitemOpen
  \bibfield  {author} {\bibinfo {author} {\bibfnamefont {David J.~C.}\
  \bibnamefont {MacKay}},\ }\href@noop {} {\emph {\bibinfo {title} {Information
  Theory, Inference \& Learning Algorithms}}}\ (\bibinfo  {publisher}
  {Cambridge University Press},\ \bibinfo {year} {2003})\BibitemShut {NoStop}%
\bibitem [{\citenamefont {Mezard}\ \emph {et~al.}(1987)\citenamefont {Mezard},
  \citenamefont {Parisi},\ and\ \citenamefont {Virasoro}}]{Mezard-book-1987}%
  \BibitemOpen
  \bibfield  {author} {\bibinfo {author} {\bibfnamefont {M.}~\bibnamefont
  {Mezard}}, \bibinfo {author} {\bibfnamefont {G.}~\bibnamefont {Parisi}}, \
  and\ \bibinfo {author} {\bibfnamefont {M.A.}\ \bibnamefont {Virasoro}},\
  }\href@noop {} {\emph {\bibinfo {title} {Spin Glass Theory and Beyond}}},\
  Lecture Notes in Physics Series\ (\bibinfo  {publisher} {World Scientific},\
  \bibinfo {year} {1987})\BibitemShut {NoStop}%
\bibitem [{\citenamefont {Mezard}\ and\ \citenamefont
  {Montanari}(2009)}]{Mezard-book-2009}%
  \BibitemOpen
  \bibfield  {author} {\bibinfo {author} {\bibfnamefont {Marc}\ \bibnamefont
  {Mezard}}\ and\ \bibinfo {author} {\bibfnamefont {Andrea}\ \bibnamefont
  {Montanari}},\ }\href@noop {} {\emph {\bibinfo {title} {Information, Physics,
  and Computation}}}\ (\bibinfo  {publisher} {Oxford University Press, Inc.},\
  \bibinfo {address} {New York, NY, USA},\ \bibinfo {year} {2009})\BibitemShut
  {NoStop}%
\bibitem [{\citenamefont {Nishimori}(2001)}]{Nishimori-book-2001}%
  \BibitemOpen
  \bibfield  {author} {\bibinfo {author} {\bibfnamefont {H.}~\bibnamefont
  {Nishimori}},\ }\href@noop {} {\emph {\bibinfo {title} {Statistical Physics
  of Spin Glasses and Information Processing: An Introduction}}},\
  International series of monographs on physics\ (\bibinfo  {publisher} {Oxford
  University Press},\ \bibinfo {year} {2001})\BibitemShut {NoStop}%
\bibitem [{\citenamefont {Perdomo-Ortiz}\ \emph {et~al.}(2016)\citenamefont
  {Perdomo-Ortiz}, \citenamefont {O'Gorman}, \citenamefont {Fluegemann},
  \citenamefont {Biswas},\ and\ \citenamefont
  {Smelyanskiy}}]{PerdomoOrtiz_SciRep2016}%
  \BibitemOpen
  \bibfield  {author} {\bibinfo {author} {\bibfnamefont {Alejandro}\
  \bibnamefont {Perdomo-Ortiz}}, \bibinfo {author} {\bibfnamefont {Bryan}\
  \bibnamefont {O'Gorman}}, \bibinfo {author} {\bibfnamefont {Joseph}\
  \bibnamefont {Fluegemann}}, \bibinfo {author} {\bibfnamefont {Rupak}\
  \bibnamefont {Biswas}}, \ and\ \bibinfo {author} {\bibfnamefont {Vadim~N.}\
  \bibnamefont {Smelyanskiy}},\ }\bibfield  {title} {\enquote {\bibinfo {title}
  {Determination and correction of persistent biases in quantum annealers},}\
  }\href@noop {} {\bibfield  {journal} {\bibinfo  {journal} {Sci. Rep.}\
  }\textbf {\bibinfo {volume} {6}},\ \bibinfo {pages} {18628} (\bibinfo {year}
  {2016})}\BibitemShut {NoStop}%
\bibitem [{\citenamefont {Mandr\`a}\ \emph {et~al.}(2017)\citenamefont
  {Mandr\`a}, \citenamefont {Zhu},\ and\ \citenamefont
  {Katzgraber}}]{Mandra2017}%
  \BibitemOpen
  \bibfield  {author} {\bibinfo {author} {\bibfnamefont {Salvatore}\
  \bibnamefont {Mandr\`a}}, \bibinfo {author} {\bibfnamefont {Zheng}\
  \bibnamefont {Zhu}}, \ and\ \bibinfo {author} {\bibfnamefont {Helmut~G.}\
  \bibnamefont {Katzgraber}},\ }\bibfield  {title} {\enquote {\bibinfo {title}
  {Exponentially biased ground-state sampling of quantum annealing machines
  with transverse-field driving hamiltonians},}\ }\href {\doibase
  10.1103/PhysRevLett.118.070502} {\bibfield  {journal} {\bibinfo  {journal}
  {Phys. Rev. Lett.}\ }\textbf {\bibinfo {volume} {118}},\ \bibinfo {pages}
  {070502} (\bibinfo {year} {2017})}\BibitemShut {NoStop}%
\bibitem [{\citenamefont {Korenkevych}\ \emph {et~al.}(2016)\citenamefont
  {Korenkevych}, \citenamefont {Xue}, \citenamefont {Bian}, \citenamefont
  {Chudak}, \citenamefont {Macready}, \citenamefont {Rolfe},\ and\
  \citenamefont {Andriyash}}]{korenkevych2016benchmarking}%
  \BibitemOpen
  \bibfield  {author} {\bibinfo {author} {\bibfnamefont {Dmytro}\ \bibnamefont
  {Korenkevych}}, \bibinfo {author} {\bibfnamefont {Yanbo}\ \bibnamefont
  {Xue}}, \bibinfo {author} {\bibfnamefont {Zhengbing}\ \bibnamefont {Bian}},
  \bibinfo {author} {\bibfnamefont {Fabian}\ \bibnamefont {Chudak}}, \bibinfo
  {author} {\bibfnamefont {William~G}\ \bibnamefont {Macready}}, \bibinfo
  {author} {\bibfnamefont {Jason}\ \bibnamefont {Rolfe}}, \ and\ \bibinfo
  {author} {\bibfnamefont {Evgeny}\ \bibnamefont {Andriyash}},\ }\bibfield
  {title} {\enquote {\bibinfo {title} {Benchmarking quantum hardware for
  training of fully visible boltzmann machines},}\ }\href@noop {} {\bibfield
  {journal} {\bibinfo  {journal} {arXiv preprint arXiv:1611.04528}\ } (\bibinfo
  {year} {2016})}\BibitemShut {NoStop}%
\bibitem [{\citenamefont {Lechner}\ \emph {et~al.}(2015)\citenamefont
  {Lechner}, \citenamefont {Hauke},\ and\ \citenamefont
  {Zoller}}]{lechner2015quantum}%
  \BibitemOpen
  \bibfield  {author} {\bibinfo {author} {\bibfnamefont {Wolfgang}\
  \bibnamefont {Lechner}}, \bibinfo {author} {\bibfnamefont {Philipp}\
  \bibnamefont {Hauke}}, \ and\ \bibinfo {author} {\bibfnamefont {Peter}\
  \bibnamefont {Zoller}},\ }\bibfield  {title} {\enquote {\bibinfo {title} {A
  quantum annealing architecture with all-to-all connectivity from local
  interactions},}\ }\href@noop {} {\bibfield  {journal} {\bibinfo  {journal}
  {Science advances}\ }\textbf {\bibinfo {volume} {1}},\ \bibinfo {pages}
  {e1500838} (\bibinfo {year} {2015})}\BibitemShut {NoStop}%
\bibitem [{\citenamefont {Realpe-G{\'o}mez}(2017)}]{realpe2017quantum}%
  \BibitemOpen
  \bibfield  {author} {\bibinfo {author} {\bibfnamefont {John}\ \bibnamefont
  {Realpe-G{\'o}mez}},\ }\bibfield  {title} {\enquote {\bibinfo {title}
  {Quantum as self-reference},}\ }\href@noop {} {\bibfield  {journal} {\bibinfo
   {journal} {arXiv preprint arXiv:1705.04307}\ } (\bibinfo {year}
  {2017})}\BibitemShut {NoStop}%
\bibitem [{\citenamefont {Choi}(2008)}]{choi2008minor}%
  \BibitemOpen
  \bibfield  {author} {\bibinfo {author} {\bibfnamefont {Vicky}\ \bibnamefont
  {Choi}},\ }\bibfield  {title} {\enquote {\bibinfo {title} {{Minor-embedding
  in adiabatic quantum computation: I. the parameter setting problem}},}\
  }\href@noop {} {\bibfield  {journal} {\bibinfo  {journal} {Quantum
  Information Processing}\ }\textbf {\bibinfo {volume} {7}},\ \bibinfo {pages}
  {193--209} (\bibinfo {year} {2008})}\BibitemShut {NoStop}%
\bibitem [{\citenamefont {Venturelli}\ \emph
  {et~al.}(2015{\natexlab{b}})\citenamefont {Venturelli}, \citenamefont
  {Mandr\`a}, \citenamefont {Knysh}, \citenamefont {O'Gorman}, \citenamefont
  {Biswas},\ and\ \citenamefont {Smelyanskiy}}]{VentuPRX}%
  \BibitemOpen
  \bibfield  {author} {\bibinfo {author} {\bibfnamefont {Davide}\ \bibnamefont
  {Venturelli}}, \bibinfo {author} {\bibfnamefont {Salvatore}\ \bibnamefont
  {Mandr\`a}}, \bibinfo {author} {\bibfnamefont {Sergey}\ \bibnamefont
  {Knysh}}, \bibinfo {author} {\bibfnamefont {Bryan}\ \bibnamefont {O'Gorman}},
  \bibinfo {author} {\bibfnamefont {Rupak}\ \bibnamefont {Biswas}}, \ and\
  \bibinfo {author} {\bibfnamefont {Vadim}\ \bibnamefont {Smelyanskiy}},\
  }\bibfield  {title} {\enquote {\bibinfo {title} {Quantum optimization of
  fully connected spin glasses},}\ }\href {\doibase 10.1103/PhysRevX.5.031040}
  {\bibfield  {journal} {\bibinfo  {journal} {Phys. Rev. X}\ }\textbf {\bibinfo
  {volume} {5}},\ \bibinfo {pages} {031040} (\bibinfo {year}
  {2015}{\natexlab{b}})}\BibitemShut {NoStop}%
\bibitem [{\citenamefont {Ackley}\ \emph {et~al.}(1985)\citenamefont {Ackley},
  \citenamefont {Hinton},\ and\ \citenamefont
  {Sejnowski}}]{ackley1985learning}%
  \BibitemOpen
  \bibfield  {author} {\bibinfo {author} {\bibfnamefont {David~H}\ \bibnamefont
  {Ackley}}, \bibinfo {author} {\bibfnamefont {Geoffrey~E}\ \bibnamefont
  {Hinton}}, \ and\ \bibinfo {author} {\bibfnamefont {Terrence~J}\ \bibnamefont
  {Sejnowski}},\ }\bibfield  {title} {\enquote {\bibinfo {title} {A learning
  algorithm for boltzmann machines},}\ }\href@noop {} {\bibfield  {journal}
  {\bibinfo  {journal} {Cognitive science}\ }\textbf {\bibinfo {volume} {9}},\
  \bibinfo {pages} {147--169} (\bibinfo {year} {1985})}\BibitemShut {NoStop}%
\bibitem [{\citenamefont {Hinton}(2012)}]{Hinton-TechRep-2012}%
  \BibitemOpen
  \bibfield  {author} {\bibinfo {author} {\bibfnamefont {Geoffrey~E.}\
  \bibnamefont {Hinton}},\ }\bibfield  {title} {\enquote {\bibinfo {title} {A
  practical guide to training restricted boltzmann machines.}}\ }in\ \href@noop
  {} {\emph {\bibinfo {booktitle} {Neural Networks: Tricks of the Trade (2nd
  ed.)}}},\ \bibinfo {series} {Lecture Notes in Computer Science}, Vol.\
  \bibinfo {volume} {7700},\ \bibinfo {editor} {edited by\ \bibinfo {editor}
  {\bibfnamefont {GrÃ©goire}\ \bibnamefont {Montavon}}, \bibinfo {editor}
  {\bibfnamefont {Genevieve~B.}\ \bibnamefont {Orr}}, \ and\ \bibinfo {editor}
  {\bibfnamefont {Klaus-Robert}\ \bibnamefont {MÃ¼ller}}}\ (\bibinfo
  {publisher} {Springer},\ \bibinfo {year} {2012})\ pp.\ \bibinfo {pages}
  {599--619}\BibitemShut {NoStop}%
\bibitem [{\citenamefont {Jaynes}(1957{\natexlab{a}})}]{jaynes1957information}%
  \BibitemOpen
  \bibfield  {author} {\bibinfo {author} {\bibfnamefont {Edwin~T}\ \bibnamefont
  {Jaynes}},\ }\bibfield  {title} {\enquote {\bibinfo {title} {Information
  theory and statistical mechanics. ii},}\ }\href@noop {} {\bibfield  {journal}
  {\bibinfo  {journal} {Physical review}\ }\textbf {\bibinfo {volume} {108}},\
  \bibinfo {pages} {171} (\bibinfo {year} {1957}{\natexlab{a}})}\BibitemShut
  {NoStop}%
\bibitem [{\citenamefont {Jaynes}(1957{\natexlab{b}})}]{Jaynes-PhysRev-1957}%
  \BibitemOpen
  \bibfield  {author} {\bibinfo {author} {\bibfnamefont {E.~T.}\ \bibnamefont
  {Jaynes}},\ }\bibfield  {title} {\enquote {\bibinfo {title} {Information
  theory and statistical mechanics},}\ }\href {\doibase
  10.1103/PhysRev.106.620} {\bibfield  {journal} {\bibinfo  {journal} {Phys.
  Rev.}\ }\textbf {\bibinfo {volume} {106}},\ \bibinfo {pages} {620--630}
  (\bibinfo {year} {1957}{\natexlab{b}})}\BibitemShut {NoStop}%
\bibitem [{\citenamefont {Jaynes}\ and\ \citenamefont
  {Bretthorst}(2003)}]{Jaynes-book-2003}%
  \BibitemOpen
  \bibfield  {author} {\bibinfo {author} {\bibfnamefont {E.T.}\ \bibnamefont
  {Jaynes}}\ and\ \bibinfo {author} {\bibfnamefont {G.L.}\ \bibnamefont
  {Bretthorst}},\ }\href@noop {} {\emph {\bibinfo {title} {Probability Theory:
  The Logic of Science}}}\ (\bibinfo  {publisher} {Cambridge University
  Press},\ \bibinfo {year} {2003})\BibitemShut {NoStop}%
\bibitem [{\citenamefont {Spall}(2003)}]{Spall-book-2003}%
  \BibitemOpen
  \bibfield  {author} {\bibinfo {author} {\bibfnamefont {James~C.}\
  \bibnamefont {Spall}},\ }\href@noop {} {\emph {\bibinfo {title} {Introduction
  to Stochastic Search and Optimization}}},\ \bibinfo {edition} {1st}\ ed.\
  (\bibinfo  {publisher} {John Wiley \& Sons, Inc.},\ \bibinfo {address} {New
  York, NY, USA},\ \bibinfo {year} {2003})\BibitemShut {NoStop}%
\bibitem [{\citenamefont {M\'ezard}\ and\ \citenamefont
  {Mora}(2009)}]{Mezard-Mora-2009}%
  \BibitemOpen
  \bibfield  {author} {\bibinfo {author} {\bibfnamefont {Marc}\ \bibnamefont
  {M\'ezard}}\ and\ \bibinfo {author} {\bibfnamefont {Thierry}\ \bibnamefont
  {Mora}},\ }\bibfield  {title} {\enquote {\bibinfo {title} {Constraint
  satisfaction problems and neural networks: A statistical physics
  perspective},}\ }\href {\doibase
  http://dx.doi.org/10.1016/j.jphysparis.2009.05.013} {\bibfield  {journal}
  {\bibinfo  {journal} {Journal of Physiology-Paris}\ }\textbf {\bibinfo
  {volume} {103}},\ \bibinfo {pages} {107 -- 113} (\bibinfo {year} {2009})},\
  \bibinfo {note} {{Neuromathematics of Vision}}\BibitemShut {NoStop}%
\bibitem [{\citenamefont {Choi}(2011)}]{choi2011minor}%
  \BibitemOpen
  \bibfield  {author} {\bibinfo {author} {\bibfnamefont {Vicky}\ \bibnamefont
  {Choi}},\ }\bibfield  {title} {\enquote {\bibinfo {title} {Minor-embedding in
  adiabatic quantum computation: Ii. minor-universal graph design},}\
  }\href@noop {} {\bibfield  {journal} {\bibinfo  {journal} {Quantum
  Information Processing}\ }\textbf {\bibinfo {volume} {10}},\ \bibinfo {pages}
  {343--353} (\bibinfo {year} {2011})}\BibitemShut {NoStop}%
\bibitem [{\citenamefont {Cai}\ \emph {et~al.}(2014)\citenamefont {Cai},
  \citenamefont {Macready},\ and\ \citenamefont {Roy}}]{cai2014practical}%
  \BibitemOpen
  \bibfield  {author} {\bibinfo {author} {\bibfnamefont {Jun}\ \bibnamefont
  {Cai}}, \bibinfo {author} {\bibfnamefont {William~G}\ \bibnamefont
  {Macready}}, \ and\ \bibinfo {author} {\bibfnamefont {Aidan}\ \bibnamefont
  {Roy}},\ }\bibfield  {title} {\enquote {\bibinfo {title} {A practical
  heuristic for finding graph minors},}\ }\href@noop {} {\bibfield  {journal}
  {\bibinfo  {journal} {arXiv:1406.2741}\ } (\bibinfo {year}
  {2014})}\BibitemShut {NoStop}%
\bibitem [{Note1()}]{Note1}%
  \BibitemOpen
  \bibinfo {note} {Blocks (a)-(d) show machine-generated pictures while blocks (e)-(h) show human-generated pictures. We have not performed the standard Turing test, where each pair of figures is shown in isolation. Ours is, in principle, a harder test for the machine as, the redundancy of having all human- and machine-generated images together enhances the probability of a human to spot differences between the two types of images. This is compensated by the low resolution of the images, which might hint at an easier test for the machine, if shown one by one, given the distortion of the images.}\BibitemShut {Stop}%
\bibitem [{\citenamefont {Kirkpatrick}\ \emph {et~al.}(1983)\citenamefont
  {Kirkpatrick}, \citenamefont {Gelatt},\ and\ \citenamefont
  {Vecchi}}]{kirkpatrick1983optimization}%
  \BibitemOpen
  \bibfield  {author} {\bibinfo {author} {\bibfnamefont {S}~\bibnamefont
  {Kirkpatrick}}, \bibinfo {author} {\bibfnamefont {CD}~\bibnamefont {Gelatt}},
  \ and\ \bibinfo {author} {\bibfnamefont {MP}~\bibnamefont {Vecchi}},\
  }\bibfield  {title} {\enquote {\bibinfo {title} {{Optimization by simulated
  annealing}},}\ }\href@noop {} {\bibfield  {journal} {\bibinfo  {journal}
  {Science}\ }\textbf {\bibinfo {volume} {220}},\ \bibinfo {pages} {671--680}
  (\bibinfo {year} {1983})}\BibitemShut {NoStop}%
\bibitem [{\citenamefont {Mastromatteo}\ and\ \citenamefont
  {Marsili}(2011)}]{Mastromatteo-JSTAT-2011}%
  \BibitemOpen
  \bibfield  {author} {\bibinfo {author} {\bibfnamefont {Iacopo}\ \bibnamefont
  {Mastromatteo}}\ and\ \bibinfo {author} {\bibfnamefont {Matteo}\ \bibnamefont
  {Marsili}},\ }\bibfield  {title} {\enquote {\bibinfo {title} {On the
  criticality of inferred models},}\ }\href@noop {} {\bibfield  {journal}
  {\bibinfo  {journal} {Journal of Statistical Mechanics: Theory and
  Experiment}\ }\textbf {\bibinfo {volume} {2011}},\ \bibinfo {pages} {P10012}
  (\bibinfo {year} {2011})}\BibitemShut {NoStop}%
\bibitem [{\citenamefont {Klymko}\ \emph {et~al.}(2012)\citenamefont {Klymko},
  \citenamefont {Sullivan},\ and\ \citenamefont {Humble}}]{Klymko2012}%
  \BibitemOpen
  \bibfield  {author} {\bibinfo {author} {\bibfnamefont {Christine}\
  \bibnamefont {Klymko}}, \bibinfo {author} {\bibfnamefont {Blair~D.}\
  \bibnamefont {Sullivan}}, \ and\ \bibinfo {author} {\bibfnamefont
  {Travis~S.}\ \bibnamefont {Humble}},\ }\bibfield  {title} {\enquote {\bibinfo
  {title} {{Adiabatic Quantum Programming: Minor Embedding With Hard
  Faults}},}\ }\href@noop {} {\bibfield  {journal} {\bibinfo  {journal}
  {arXiv:1210.8395}\ } (\bibinfo {year} {2012})}\BibitemShut {NoStop}%
\bibitem [{\citenamefont {Job}\ and\ \citenamefont {Lidar}(2017)}]{Job2017}%
  \BibitemOpen
  \bibfield  {author} {\bibinfo {author} {\bibfnamefont {Joshua}\ \bibnamefont
  {Job}}\ and\ \bibinfo {author} {\bibfnamefont {Daniel}\ \bibnamefont
  {Lidar}},\ }\bibfield  {title} {\enquote {\bibinfo {title} {Test-driving 1000
  qubits},}\ }\href@noop {} {\bibfield  {journal} {\bibinfo  {journal}
  {arXiv:1706.07124}\ } (\bibinfo {year} {2017})}\BibitemShut {NoStop}%
\bibitem [{\citenamefont {Katzgraber}(2017)}]{Katzgraber2017}%
  \BibitemOpen
  \bibfield  {author} {\bibinfo {author} {\bibfnamefont {Helmut~G.}\
  \bibnamefont {Katzgraber}},\ }\bibfield  {title} {\enquote {\bibinfo {title}
  {Viewing vanilla quantum annealing through spin glasses},}\ }\href@noop {}
  {\bibfield  {journal} {\bibinfo  {journal} {arXiv:1708.08885}\ } (\bibinfo
  {year} {2017})}\BibitemShut {NoStop}%
\bibitem [{\citenamefont {R{\o}nnow}\ \emph {et~al.}(2014)\citenamefont
  {R{\o}nnow}, \citenamefont {Wang}, \citenamefont {Job}, \citenamefont
  {Boixo}, \citenamefont {Isakov}, \citenamefont {Wecker}, \citenamefont
  {Martinis}, \citenamefont {Lidar},\ and\ \citenamefont
  {Troyer}}]{ronnow2014defining}%
  \BibitemOpen
  \bibfield  {author} {\bibinfo {author} {\bibfnamefont {Troels~F}\
  \bibnamefont {R{\o}nnow}}, \bibinfo {author} {\bibfnamefont {Zhihui}\
  \bibnamefont {Wang}}, \bibinfo {author} {\bibfnamefont {Joshua}\ \bibnamefont
  {Job}}, \bibinfo {author} {\bibfnamefont {Sergio}\ \bibnamefont {Boixo}},
  \bibinfo {author} {\bibfnamefont {Sergei~V}\ \bibnamefont {Isakov}}, \bibinfo
  {author} {\bibfnamefont {David}\ \bibnamefont {Wecker}}, \bibinfo {author}
  {\bibfnamefont {John~M}\ \bibnamefont {Martinis}}, \bibinfo {author}
  {\bibfnamefont {Daniel~A}\ \bibnamefont {Lidar}}, \ and\ \bibinfo {author}
  {\bibfnamefont {Matthias}\ \bibnamefont {Troyer}},\ }\bibfield  {title}
  {\enquote {\bibinfo {title} {Defining and detecting quantum speedup},}\
  }\href@noop {} {\bibfield  {journal} {\bibinfo  {journal} {Science}\ }\textbf
  {\bibinfo {volume} {345}},\ \bibinfo {pages} {420--424} (\bibinfo {year}
  {2014})}\BibitemShut {NoStop}%
\end{thebibliography}
\end{document}